\documentclass[11pt]{article}
\makeatletter

\newif\if@alec@long

\RequirePackage{iftex}
\ifPDFTeX
    \RequirePackage[utf8]{inputenc} % UTF-8 input is the default on XeLaTeX and LuaLaTeX
\fi
\@ifclassloaded{acmart}{}{}
\RequirePackage{mathtools}
\@ifpackageloaded{amsthm}{}{\@ifpackageloaded{newtxmath}{}{}}
\RequirePackage{amsfonts, amsthm, arcs, bbm, bm, booktabs, caption, CJKutf8, csquotes, enumitem, epigraph, float, geometry, graphicx, hyperref, mathdots, mathrsfs, multirow, pgfplots, physics, shuffle, subcaption, tabularx, tcolorbox, tikz, tikz-cd, todonotes}
\hypersetup{colorlinks=true,citecolor=blue,urlcolor=gray}
\RequirePackage[capitalise]{cleveref}
\crefname{figure}{Figure}{Figures}
\crefname{algocf}{Algorithm}{Algorithms}
\@ifpackageloaded{natbib}{}{\RequirePackage[numbers]{natbib}}
\pgfplotsset{compat=1.16}

\@ifpackageloaded{newtxmath}{}{\RequirePackage{amssymb}}

\RequirePackage[ruled]{algorithm2e}

\@ifclassloaded{acmart}{%
    
    \SetAlFnt{\small}
    \SetAlCapFnt{\small}
    \SetAlCapNameFnt{\small}
    \SetAlCapHSkip{0pt}
    \IncMargin{-\parindent}
}{%
    
    \@ifundefined{oct}{}{} % newtxmath defines \oct itself
    \@ifclassloaded{beamer}{}{\geometry{margin=1in}}
}

\@ifclassloaded{beamer}{
    \usetheme{Frankfurt}
    \usecolortheme{beaver}
    
    \setlength{\marginparwidth}{1in} 
    \presetkeys{todonotes}{inline}{}
}{
    
    \newtheorem{thm}{Theorem}
    \newtheorem{theorem}[thm]{Theorem}
    
    \newtheorem{lemma}[thm]{Lemma}
    \newtheorem{proposition}[thm]{Proposition}

    \newtheorem{remark}[thm]{Remark}

    \newtheorem{example}[thm]{Example}

    \crefname{thm}{Theorem}{Theorems}
    \@for\alec@thm:={theorem,corollary,lemma,proposition,definition,problem,conjecture,claim,fact,remark,note,question,solution,example,assumption}\do{%
        \edef\alec@hook{\noexpand\AddToHook{env/\alec@thm/begin}{\noexpand\crefalias{thm}{\alec@thm}}}%
        \alec@hook
    }
    \crefname{theorem}{Theorem}{Theorems}
    \crefname{corollary}{Corollary}{Corollaries}
    \crefname{lemma}{Lemma}{Lemmas}
    \crefname{proposition}{Proposition}{Propositions}
    \crefname{definition}{Definition}{Definitions}
    \crefname{problem}{Problem}{Problems}
    \crefname{conjecture}{Conjecture}{Conjectures}
    \crefname{claim}{Claim}{Claims}
    \crefname{fact}{Fact}{Facts}
    \crefname{remark}{Remark}{Remarks}
    \crefname{note}{Note}{Notes}
    \crefname{question}{Question}{Questions}
    \crefname{solution}{Solution}{Solutions}
    \crefname{example}{Example}{Examples}
    \crefname{assumption}{Assumption}{Assumptions}

    \newtheorem*{definition*}{Definition}
    \newtheorem*{theorem*}{Theorem}
    \newtheorem*{claim*}{Claim}
    \newtheorem*{example*}{Example}
    \newtheorem*{question*}{Question}
    \newtheorem*{problem*}{Problem}
    \newtheorem*{exercise*}{Exercise}
    \newtheorem*{solution*}{Solution}
    \newtheorem*{remark*}{Remark}
    \newtheorem*{note*}{Note}
    \newtheorem*{assumption*}{Assumption}
}

\@ifclassloaded{amsart}{%
    
    \let\oldtocsection=\tocsection
    \let\oldtocsubsection=\tocsubsection
    \renewcommand{\tocsection}[2]{\hspace{0em}\oldtocsection{#1}{#2}}
    \renewcommand{\tocsubsection}[2]{\hspace{1em}\oldtocsubsection{#1}{#2}}
    
}{%

}

\if@alec@long
    \numberwithin{equation}{section}
    \@ifundefined{c@thm}{}{\numberwithin{thm}{section}}
\fi

\newcommand{\te}{\mathrm}
\newcommand{\tbf}{\textbf}

\newcommand{\cd}{\cdot}
\newcommand{\cds}{\cdots}
\newcommand{\lds}{\ldots}

\newcommand{\fr}{\frac}

\newcommand{\teif}{\mathrm{if}\,}

\DeclareMathOperator{\sgn}{sign}

\DeclareMathOperator*{\argmax}{argmax}

\newcommand{\bE}{\mathbb E}

\newcommand{\bP}{\mathbb P}

\newcommand{\bR}{\mathbb R}

\newcommand{\one}{\mathbf{1}}

\newcommand{\cY}{\mathcal Y}

\newcommand{\bp}[1]{\left( #1 \right)} % (x)
\newcommand{\bb}[1]{\left[ #1 \right]} % [x]
\newcommand{\bc}[1]{\left\{ #1 \right\}} % {x}
\newcommand{\case}[1]{\begin{cases} #1 \end{cases}}

\newcommand{\Binom}[1]{\operatorname{Binom} \mathopen{} \left( #1 \right)}
\newcommand{\Bern}[1]{\operatorname{Bern} \mathopen{} \left( #1 \right)}

\newcommand{\E}[1]{\bE \mathopen{} \left[ #1 \right]}
\renewcommand{\P}[1]{\bP \mathopen{} \left[ #1 \right]}

\newcommand{\pd}[2]{\frac{\partial #1}{\partial #2}}

\makeatother
\title{Two is better than one: optimizing binary rating scales for online platforms}

\author{Liren Shan\thanks{Authors are listed in alphabetical order.} \\ \small Toyota Technological Institute at Chicago \and Alec Sun\footnotemark[1] \\ \small University of Chicago \and Jingyan Wang\footnotemark[1] \\ \small Toyota Technological Institute at Chicago}

\date{}

\begin{document}

\maketitle

\begin{abstract}
    Online platforms often struggle with inflated, uninformative user ratings. This issue can be mitigated by better design of the data collection mechanism, for example through changing the phrasing of user questions. Previous work optimizes a single rating scale to best distinguish the inherent qualities of items under binary feedback. In this paper, we demonstrate the power of utilizing different rating scales for different users. First, we show that a designer can always split a single rating scale into two to improve item rankings. Next, we formally characterize the optimal design of multiple scales as a series of 0-1 step functions whose threshold values are evenly spaced. Finally, we consider an adaptive model where subsequent rating scales can depend on the responses from previous users and items are ranked by their full sequence of responses, and show that adaptivity leads to exponentially greater distinguishing power. All in all, our theoretical results show that using multiple rating scales significantly improves item rankings on platforms.

\end{abstract}

\section{Introduction} \label{s:intro}

User data is the backbone of online platforms. Ratings are collected, for example, for videos on YouTube, products on Amazon, businesses on Yelp, freelancers on Taskrabbit, and so on. Such rating data provides valuable information to users to inform their decision making \citep{rachmiani2024}. Moreover, recommendation and personalization algorithms rely on user data to estimate the underlying quality of products, learn user preferences, and make future recommendations. However, ratings are often uninformative in practice, largely due to \emph{rating inflation}, a tendency for users to give very high ratings to most items \citep{aziz2023consequences}. For example, 74\% of guest reviews for Airbnb properties receive a perfect rating of 5 stars \citep{fradkin2021}. The percentage of positive ratings for an eBay seller has a median of 100\% and a mean of $99.3\%$ \citep{nosko2015}. Under inflation, it becomes difficult to distinguish the quality of items. Users need to rely on small differences that are often noisy in nature, for example, comparing restaurants with scores of 4.8 and 4.9. Beyond inflation, another common pattern for user ratings is a bimodal shape, where ratings tend to be either very high or very low to convey that users were either very satisfied or very dissatisfied. These ratings again become noisy and uninformative.

Prior work such as \citet{garg2019} has made the observation that the distribution of user ratings is not fixed, but instead depends on the design of data elicitation. The language used in asking a question influences the user responses. For example, for binary questions, let us compare the two questions ``did you have a positive experience?'' and ``was your experience the best you have ever had?''. It is natural to anticipate, and empirically validated \citep{garg2019}, that the former question induces users to apply a more lenient calibration and provide more positive ratings. By designing these questions carefully, the platform mitigates rating inflation, increases data quality, makes more effective future recommendations, and thus improves the user experience on the platform.

We follow \citet{garg2019} and consider binary responses because they are simple to implement and consistent with the well-known fact that people are unable to use fine-grained numerical values to indicate subtle differences accurately \citep{miller1956seven}. The design of binary rating scales is mathematically formalized as follows. We represent a rating scale by a nondecreasing function $\beta:[0, 1]\rightarrow [0, 1]$, mapping the item's inherent quality to the likelihood that a user provides a positive rating. Specifically, an item with true quality $\theta\in [0, 1]$ receives a positive rating from a user with probability $\beta(\theta)$. Going back to the previous example questions, ``did you have a positive experience?'' may be represented by $\beta(\theta) \approx \theta$, where the probability of an item being rated positively smoothly increases in its quality $\theta$. In contrast, the question ``was your experience the best you have ever had?'' corresponds to a function $\beta$ whose value is close to $0$ unless $\theta$ is close to 1.

Under this model, the goal is to compute the optimal binary scale $\beta$ under a certain objective, to be introduced formally in \cref{s:model}, for distinguishing the quality of items. For a single binary rating scale, \citet{garg2019} construct an optimization problem that can be solved numerically to compute the scale that maximizes the speed at which items of different qualities can be distinguished from each other. This optimal scale is then used to elicit binary ratings from all users on the platform.

Departing from prior work, in this paper we are motivated by the fact that it is feasible and convenient to apply different rating scales to different users, for example by asking half of the users if their experience was above average and the other half if their experience was the best they have ever had. Allowing multiple rating scales generalizes a single rating scale because one can always set all the scales to be the same. Hence using multiple scales weakly dominates using a single scale in distinguishing items, but the precise gains are a priori unclear. Our research question is therefore:

\begin{quote}
    \emph{How should a platform design multiple scales, and what is the quantitative gain  compared to a single scale?}
\end{quote}

\paragraph{Overview of results.}

 We characterize the gain from using multiple rating scales for data collection through the following theoretical results:
\begin{itemize}
    \item Given a single rating scale applied to all users, one can always construct a pair of rating scales, where each rating scale is applied to half of the users, such that the ranking objective under the two rating scales is at least that under the single scale, and strictly higher unless the single scale is already a 0-1 step function or a constant.

    \item If we allow the number of rating scales to equal the number of users, so that each user rates on a different scale, then we derive the optimal solution exactly: each user rates under a distinct 0-1 step function $f_\tau(\theta) = \case{0 & \teif \theta < \tau \\ 1 & \teif \theta \ge \tau}$ with threshold at $\tau$.

    \item Under an \emph{adaptive} model where users arrive sequentially, we consider scales chosen adaptively based on the rating scales and responses of all previous users, and we let the platform rank items by their full sequence of responses rather than by their number of positive ratings. We again derive the optimal solution, where each rating scale is still a 0-1 step function but with different threshold values compared to the non-adaptive setting: the scales perform a binary search for the item's quality. We show that adaptivity provides an exponential gain in the objective compared to the non-adaptive setting, and that this gain requires changing the aggregation rule in addition to the elicitation rule.
\end{itemize}

% \todo{How much better does $k$ scales perform than one scale? Also is there an efficient algorithm for computing the best single scale for $k$ ratings?}

\section{Related work}

\paragraph{Rating inflation and rating system design.}

Ratings in online marketplaces often become inflated and uninformative, due to the pressure of not hurting sellers \citep{filippas2022inflation}, reciprocity between sellers and buyers \citep{dellarocas2008silence}, and so on. The design of the rating system, through changing the verbal description of each point on the scale \citep{klockars1988positions,garg2021labor}, has been proposed as an effective way to mitigate inflation. More quantitatively, \citet{garg2019} compute the binary rating scale that optimizes certain measures of rating informativeness.

\paragraph{Miscalibration.}
Ratings from people are known to be miscalibrated \citep{griffin2004judgment}, and various computation methods are proposed to correct such miscalibration in a post-hoc fashion \citep{paul1981examiners,roos2011quadratic,mackay2017assessment}. Instead of correcting the scores collected under the inherent miscalibration of people, the design of rating systems can be viewed as changing this calibration as part of the data collection process.

\paragraph{Aggregating heterogeneous data in crowdsourcing.}
Our subsequent results demonstrate the gain from using heterogeneous scales. At a high level, the gain comes from each scale specializing in part of the information collection task. For example, one scale is effective in distinguishing items of low quality, and another scale is effective in distinguishing top items. This specialization is similar in spirit to aggregating heterogeneous human data in crowdsourcing \citep{dawid1979em,welinder2010multidimensional}. Again, instead of learning the heterogeneous expertise of people from data, the rating scale allows us to endogenously induce the heterogeneity under which information collection is the most effective.

\paragraph{Item response theory.}
Item response theory \citep{embretson2013irt} in psychometrics concerns estimating the latent quality of students based on their responses to test questions, which is similar to our problem of ranking items based on user ratings. Our adaptive setting is also similar to adaptive testing, where test questions are selected adaptively based on a student's previous responses. Instead of choosing from a fixed bank of questions or rating scales, we characterize the optimal ``response functions'' to guide the design of questions.

\section{Problem formulation} \label{s:model}

%\paragraph{Item qualities.} 
The platform wishes to rank a set of items where each item has a unique and unknown quality $\theta\in [0, 1]$. We assume without loss of generality that these qualities are uniformly distributed in $[0, 1]$, noting that we can reduce any distribution to this case by rescaling, so that $\theta$ is the item's quality \emph{quantile} rather than its absolute quality. We assume that each item will accumulate a total of $k$ ratings from users on the platform.

\paragraph{Rating scales.} Each user's rating is binary and collected under a \emph{rating scale}, which is a nondecreasing function $\beta(\theta):[0,1]\to [0,1]$ such that an item with quality $\theta$ gets a positive rating with probability $\beta(\theta)$ and a negative rating with probability $1 - \beta(\theta)$. Ratings of different items and of different users are independent given the item qualities. Prior work \citep{garg2019} has primarily been concerned with optimizing a \emph{single} rating scale $\beta$ that distinguishes items of different qualities. For a single scale, ratings for an item of quality $\theta$ are always distributed as $\Bern{\beta(\theta)}$ and hence the number of positive ratings after $k$ users is $\Binom{k, \beta(\theta)}$. Moving beyond a single scale, we consider \emph{multiple scales}, where the rating scales $\beta(\theta)$ need not be identical across the $k$ users. For each $i\in [k]$, we denote the rating scale used for the $i$-th user by $\beta_i$. The total number of positive ratings that an item with quality $\theta$ accumulates is a random variable $n_k(\theta) := \sum_{i=1}^k \Bern{\beta_i(\theta)}$.

\paragraph{Platform objective.}
The platform aims to design a mechanism for collecting user ratings such that the ranking induced by the mean ratings of these items is as accurate as possible. That is, given two items with qualities $\theta_1 < \theta_2$, the platform ranks them accurately if $\theta_2$ accumulates more positive ratings than $\theta_1$, namely $n_k(\theta_1) < n_k(\theta_2)$. We consider an accuracy metric whose value is $1$ if the two items are ranked correctly, $-1$ if they are ranked incorrectly, and $0$ if they are tied in terms of their mean ratings. This ranking accuracy between items with qualities $\theta_1 < \theta_2$ is thus captured by the expression:
\begin{align*}
    P_k(\theta_1, \theta_2\mid \mathbf{\beta}) := \P{n_k(\theta_1) < n_k(\theta_2)} - \P{n_k(\theta_1) > n_k(\theta_2)}.
\end{align*}
The platform generally wishes to maximize $P_k(\theta_1, \theta_2\mid \mathbf{\beta})$ across all pairs of item qualities $\theta_1 < \theta_2$. However, it may care about ranking certain pairs of qualities more than others, which is captured by an integrable \emph{weight function} $w(\theta_1, \theta_2):[0,1]\times [0,1]\to \bR_{\ge 0}$ that indicates how much the platform values ranking items with qualities $\theta_1 < \theta_2$ correctly.

\begin{example}
    The most commonly used weight functions are the following:
    \begin{itemize}
        \item Kendall's $\tau$ function weights each pair of qualities uniformly, namely $w(\theta_1, \theta_2) = 1$ for all $\theta_1, \theta_2\in [0, 1]$.
    
        \item Spearman's $\rho$ function weights pairs of qualities proportional to the difference in qualities, namely $w(\theta_1, \theta_2) = \theta_2 - \theta_1$ for all $\theta_1, \theta_2\in [0,1]$. In words, it is more important to distinguish items with significantly different qualities than those with similar qualities.
    \end{itemize}
    More generally, varying the weight function $w(\theta_1, \theta_2)$ allows the platform to encode a rich set of objectives \citep{tarsitano2009, costa2006}.

    % For example, the following weight function captures a platform's goal to distinguish the top half of items from the bottom half, but does not care about ranking items within these two quantiles may use the  $$w(\theta_1, \theta_2) = \case{1 & \teif \bp{\fr12 - \theta_1}\bp{\fr12 - \theta_2} < 0 \\ 0 & \text{otherwise}.}$$ 
\end{example}

The platform seeks to maximize the following overall objective, which is obtained by integrating $P_k(\theta_1, \theta_2\mid \mathbf{\beta})$ across all $\theta_1 < \theta_2$, weighting each pair of qualities $(\theta_1, \theta_2)$ by $w(\theta_1, \theta_2)$, and finally normalizing by a factor of 2 since $\P{\theta_1 < \theta_2} = \fr12$:

%\begin{tcolorbox}[title=Platform's objective function]
    $$W_k := 2\int_{\theta_1 < \theta_2} w(\theta_1, \theta_2) \cd P_k(\theta_1, \theta_2\mid \mathbf{\beta})\,d\theta_1 d\theta_2.$$ %where $$P_k(\theta_1, \theta_2\mid \mathbf{\beta}) := \P{n_k(\theta_1) < n_k(\theta_2)} - \P{n_k(\theta_1) > n_k(\theta_2)}.$$
%\end{tcolorbox}

%

Note that the theoretical maximum of $W_k$ under Kendall's $\tau$ function is 1, which occurs when all items can be perfectly distinguished from each other, namely $P_k(\theta_1, \theta_2\mid \mathbf{\beta}) = 1$ for all $\theta_1 < \theta_2$.

\paragraph{Comparing single and multiple scales.} Recall that the goal of this paper is to understand the power of using different rating scales $\{\beta_i\}_{i=1}^k$ for different users rather than using a single scale $\beta$ for all $k$ ratings. To do this, we would like to compute the maximum platform objective allowing multiple rating scales, which we denote by $W^\te{multiple}_k$, and show that it is greater than the maximum objective that uses a single rating scale, which we denote by $W^\te{single}_k$. For this comparison to be meaningful, we must use the same weight function $w(\theta_1, \theta_2)$, so we use Kendall's $\tau$ function for ease of exposition. Our theoretical results also show how to generalize to other weight functions $w(\theta_1, \theta_2)$. In \cref{s:multiple} we prove that $W^\te{multiple}_k = 1 - \fr{1}{k+1}$ for Kendall's $\tau$ function if each rating can be on its own scale, which means that using multiple scales approaches perfect accuracy at a rate of order $\frac{1}{k}$. On the other hand, we are unable to compute $W^\te{single}_k$ exactly. Instead, we employ two methods to \emph{estimate} $W^\te{single}_k$. The first method is deriving a theoretical upper bound $\bar{W}^\te{single}_k$ for $W^\te{single}_k$:

\begin{proposition}[Upper bound on platform objective for single scale] \label{thm:single-value}
Let $w(\theta_1, \theta_2)$ be Kendall's $\tau$ function. The best platform objective $W^\te{single}_k$ for $k$ ratings using a single scale is at most
\begin{align}\label{eq:ub_single_scale}
    \bar{W}^\te{single}_k :=1-2^{k-1}\int_0^1 \frac{y^k}{(1+y)^{k-1}}\,dy .
\end{align}
\end{proposition}

The proof of \cref{thm:single-value} is in \Cref{a:single}. The second method is solving for the best \emph{discretized} single scale. To do this, we split $[0,1]$ into $n$ equally spaced intervals and solve for the optimal step function that is constant within each interval and maximizes $W_k$. The results of this computation are shown in \Cref{fig:single-scale-step}. 

\paragraph{Adaptive rating scales.} The model above, following \cite{garg2019}, has two components. The first is an \emph{elicitation} rule: the $k$ scales $\beta_1,\lds,\beta_k$ are set by the designer in advance, so the scale shown to a user does not depend on the ratings of previous users. The second is an \emph{aggregation} rule: an item's score is its number of positive ratings $n_k(\theta)$, and the item with the higher count is ranked higher. In \Cref{s:adaptive} we also consider an \emph{adaptive} model that modifies both components. For elicitation, the designer uses a scale $\beta_1$ for the first user, then uses the first user's binary rating to choose the second scale $\beta_2$, and in general chooses the scale $\beta_j$ for the $j$-th user as a function of the ratings of the first $j-1$ users. For aggregation, the platform ranks items by their full sequence of $k$ binary responses rather than by the number of positive responses. This sequence identifies the path that the item took through the tree of scales. We show in \Cref{s:adaptive} that the second change is necessary for the first to be useful, and that the optimal configuration of adaptive rating scales distinguishes items at an exponentially better rate of $W^\te{adaptive}_k = 1-\frac{1}{2^k}$ compared to the non-adaptive rate of $W^\te{multiple}_k = 1-\frac{1}{k+1}$.

\section{Local result: two is better than one} \label{s:splitting}

We first show a \emph{local} improvement result: Starting with any single scale $\beta$ used for an even number of ratings, we construct two scales $\beta^+$ and $\beta^-$, such that using each scale for half of the ratings improves the platform objective compared to using $\beta$ for all ratings.

\begin{theorem} \label{thm:splitting}
    Let $k = 2m$ be any even integer, and let $w$ be any nonnegative weight function. 
    Given any single rating scale $\beta(\theta)$ for $k$ ratings, there exist two scales
    \[
        \beta^+(\theta):=\min\{2\beta(\theta),1\},
        \qquad
        \beta^-(\theta):=\max\{2\beta(\theta)-1,0\}
    \] such that the platform objective $W_k$ achieved by using the scale $\beta^+$ for $m$ ratings and the scale $\beta^-$ for the other $m$ ratings is at least that achieved by using scale $\beta$ for all $k$ ratings. Furthermore, if $w>0$ almost everywhere, the increase in $W_k$ is strict unless $\beta$ is a 0-1 step function or a constant.
\end{theorem}

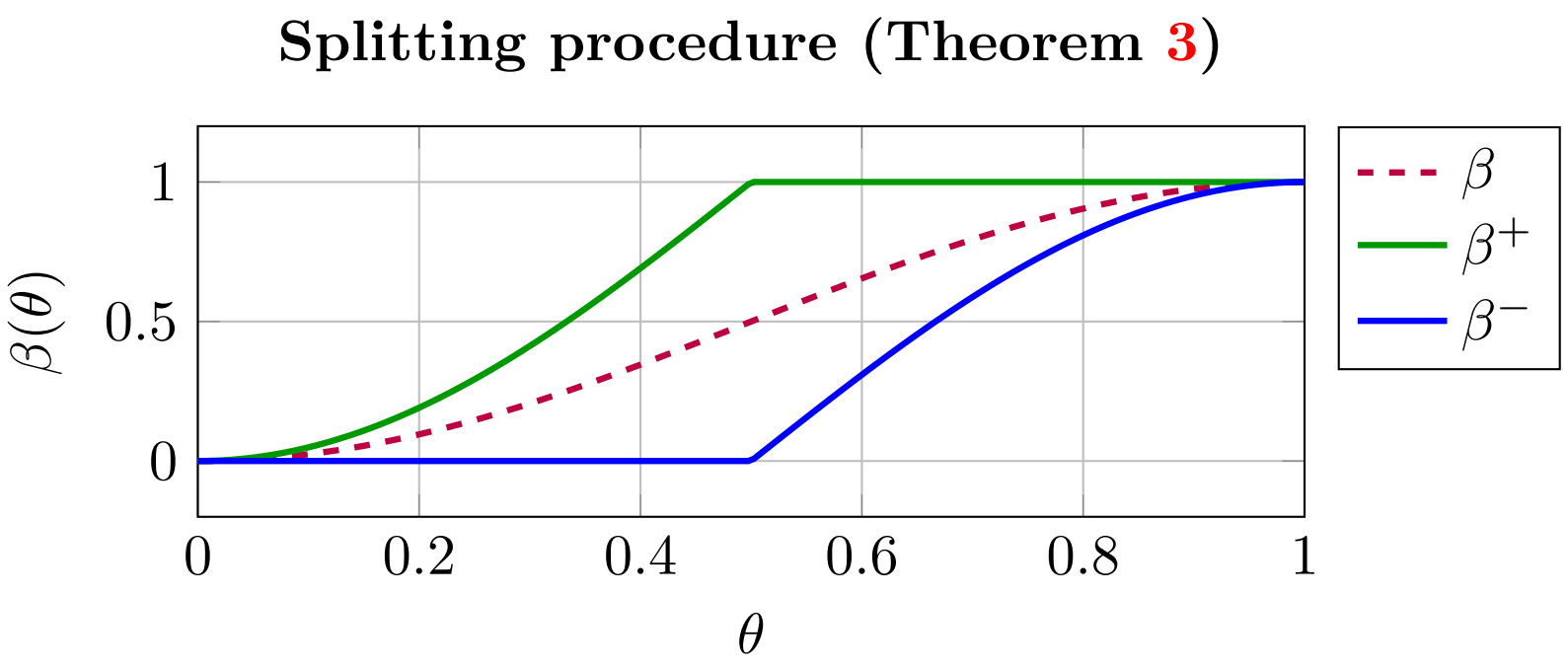
\begin{figure}
\centering
\begin{tikzpicture}[
    % Declare the functions explicitly
    declare function={
        f(\x) = 0.5 * (1 + sin(180*\x - 90));
        g(\x) = min(2*f(\x), 1);
        h(\x) = max(2*f(\x) - 1, 0);
    }
]
    \begin{axis}[
        domain=0:1,
        samples=200,          % High sample rate for smooth curves
        xmin=0, xmax=1,
        ymin=-0.2, ymax=1.2, % Slight padding on y-axis for visibility
        xlabel={$\theta$},
        ylabel={$\beta(\theta)$},
        grid=major,
        legend pos=outer north east, 
        legend cell align={left},
        width=9cm,
        height=4.2cm,
        title={\tbf{Splitting procedure (\cref{thm:splitting})}}
    ]

        % Plot f(x)
        \addplot[purple, dashed, very thick] {f(x)};
        \addlegendentry{$\beta$}

        % Plot g(x)
        \addplot[green!60!black, solid, very thick] {g(x)};
        \addlegendentry{$\beta^+$}

        % Plot h(x)
        \addplot[blue, solid, very thick] {h(x)};
        \addlegendentry{$\beta^-$}

    \end{axis}
\end{tikzpicture}
\caption{Splitting a single scale $\beta(\theta)$ into two scales $\beta^+(\theta) = \min\bc{2\beta(\theta), 1}$ and $\beta^-(\theta) = \max\bc{2\beta(\theta)-1, 0}$ increases the platform objective $W_k$.\label{fig:split}}
\end{figure}
The scales $\beta^+$ and $\beta^-$ are constructed by ``splitting'' the original scale $\beta$, visualized in \Cref{fig:split}. 
Intuitively, a steep curve in the scale suggests that users respond very differently to a pair of items, and hence the rating scale achieves a higher accuracy. In this construction, the scale $\beta^+$ focuses on distinguishing items of low quality, by amplifying the original scale $\beta$ for these items. Likewise, the scale $\beta^-$ focuses on distinguishing items of high quality. 
This splitting preserves the expected number of positive ratings while reducing the noise in the ratings. 
For example, consider $\theta$ with $p := \beta(\theta) < 1/2$. The two split ratings are distributed as $\Bern{2p}$ and $0$, whose total variance is $2p(1-2p)$, smaller than the variance $2p(1-p)$ of two independent $\Bern{p}$ ratings. 
% \jw{I think one intuitive explanation is as follows. Consider Bernoulli(p). Under a single scale, it's the sum of two Bern(p) with variance $2\cdot p(1-p)$. Under the new scale, it's one rv Bern(2p), with variance $2p\cdot (1-2p)$ which is smaller than former.}

We also note that the splitting procedure can be performed recursively to further increase the platform objective. Starting with any number of scales, we split any one of these scales $\beta$ used for $k = 2m$ ratings into two scales each used for $m$ ratings, and \cref{thm:splitting} guarantees that the platform objective increases with every split. In particular, this means that allowing for one more scale is always better.

The full proof of \cref{thm:splitting} is technically involved and can be found in \cref{a:splitting}. Here we provide an outline. Let $p:=\beta(\theta_1) < q:=\beta(\theta_2)$. The pairwise objective $P_k$ for using $\beta$ on all $k = 2m$ ratings is \[P_k(\theta_1, \theta_2\mid \beta) = K_{2m}(p,q) := \E{\sgn\!\bp{\Binom{2m, q}- \Binom{2m, p}}}\] where $\sgn(x) = \case{1 & \teif x > 0 \\ 0 & \teif x = 0 \\ -1 & \teif x < 0}$ is the sign function. To formalize the split, we define the bounding functions $B^+(x):=\min\{2x,1\}$ and $B^-(x):=\max\{2x-1,0\}$ so that \(\beta^+= B^+ \circ \beta\) and \(\beta^- = B^- \circ \beta\). An item with quality $\theta$ and $p = \beta(\theta)$ has $Y_p \sim \Binom{m,B^+(p)}+\Binom{m, B^-(p)}$ positive ratings, hence the pairwise objective for using $\beta^+$ and $\beta^-$ on $m$ ratings each is
\[
P_k(\theta_1, \theta_2\mid \beta^+, \beta^-) = \widetilde K_{2m}(p,q)
:=
\E{\sgn\!\bp{Y_q-Y_p}}.
\]
To prove \cref{thm:splitting} it suffices to show that $\widetilde K_{2m}(p,q)\ge K_{2m}(p,q)$ for all $p < q$, as the result follows by taking a weighted average over $\theta_1 < \theta_2$ according to the weight function $w(\theta_1, \theta_2)$. In \cref{a:splitting} we show that $\widetilde K_{2m}(p,q)\ge K_{2m}(p,q)$ via a sequence of rating hybrids: we start with the original \(k = 2m\) copies of $\beta$-scales and replace a pair of $\beta$-scales by one $\beta^+$-scale and one $\beta^-$-scale, one pair at a time. The key claim is that each such replacement weakly increases the objective $W_k$, and strictly increases the pairwise objective for every pair with $\beta(\theta_1)<\beta(\theta_2)$ other than $(\beta(\theta_1),\beta(\theta_2))=(0,1)$. This is what makes the gain strict unless $\beta$ is a 0-1 step function or a constant. A constant scale has objective $0$ before and after splitting.

\section{Global result: deriving the optimal multiple scales} \label{s:multiple}

%%%
\subsection{One rating per scale}

In this section, we derive the optimal configuration of scales when the platform may choose $k$ possibly different rating scales, one for each of the $k$ ratings, so that each scale is used for exactly one rating.

\begin{theorem}\label{thm:multiple}
    Consider any weight function $w(\theta_1, \theta_2)$. Given $k$ scales $\beta_1,\lds,\beta_k$ to be used for $k$ ratings, where the $i$-th rating uses scale $\beta_i$, 
    the maximum platform objective is achieved by 0-1 step functions $\beta_i^*(\theta) = \one\bb{\theta \ge \tau_i}$ for all $i\in [k]$, with threshold values $0 \le \tau_1\le \cdots \le \tau_k \le 1$ that depend on $w(\theta_1, \theta_2)$. The thresholds are distinct and lie in $(0,1)$ whenever $w>0$ almost everywhere. When $w(\theta_1,\theta_2)$ is Kendall's $\tau$ function, the threshold values are evenly spaced in $[0,1]$, that is, 
   $\tau_i =\frac{i}{k+1}$ for all $ i\in [k]$.
\end{theorem}

\begin{figure}
\centering
    \begin{tikzpicture}
        \begin{axis}[
            width=9cm, height=4.2cm,
            domain=0:1,
            xmin=0, xmax=1,
            ymin=-0.2, ymax=1.2,
            xlabel={$\theta$},
            ylabel={$\beta(\theta)$},
            title={\tbf{Optimal configuration of $k$ scales (\cref{thm:multiple})}},
            xtick={0, 0.2, 0.4, 0.6, 0.8, 1},
            ytick={0, 0.5, 1},
            grid=major,
            grid style={dashed},
            legend pos=outer north east,
            samples=100
        ]
        
        % Threshold 1/5
        \addplot[color=red, very thick] coordinates {(0,0) (0.2,0) (0.2,1) (1,1)};
        \addlegendentry{$\beta_1$}

        % Threshold 2/5
        \addplot[color=orange, very thick] coordinates {(0,0) (0.4,0) (0.4,1) (1,1)};
        \addlegendentry{$\beta_2$}

        % Threshold 3/5
        \addplot[color=green!70!black, very thick] coordinates {(0,0) (0.6,0) (0.6,1) (1,1)};
        \addlegendentry{$\beta_3$}

        % Threshold 4/5
        \addplot[color=blue, very thick] coordinates {(0,0) (0.8,0) (0.8,1) (1,1)};
        \addlegendentry{$\beta_4$}
        
        \end{axis}
    \end{tikzpicture}

\caption{For Kendall's $\tau$ function, the optimal $k$ scales for $k$ ratings are 0-1 step functions with evenly spaced thresholds in $[0,1]$. The figure illustrates the optimal configuration of $k=4$ scales.}
\end{figure}
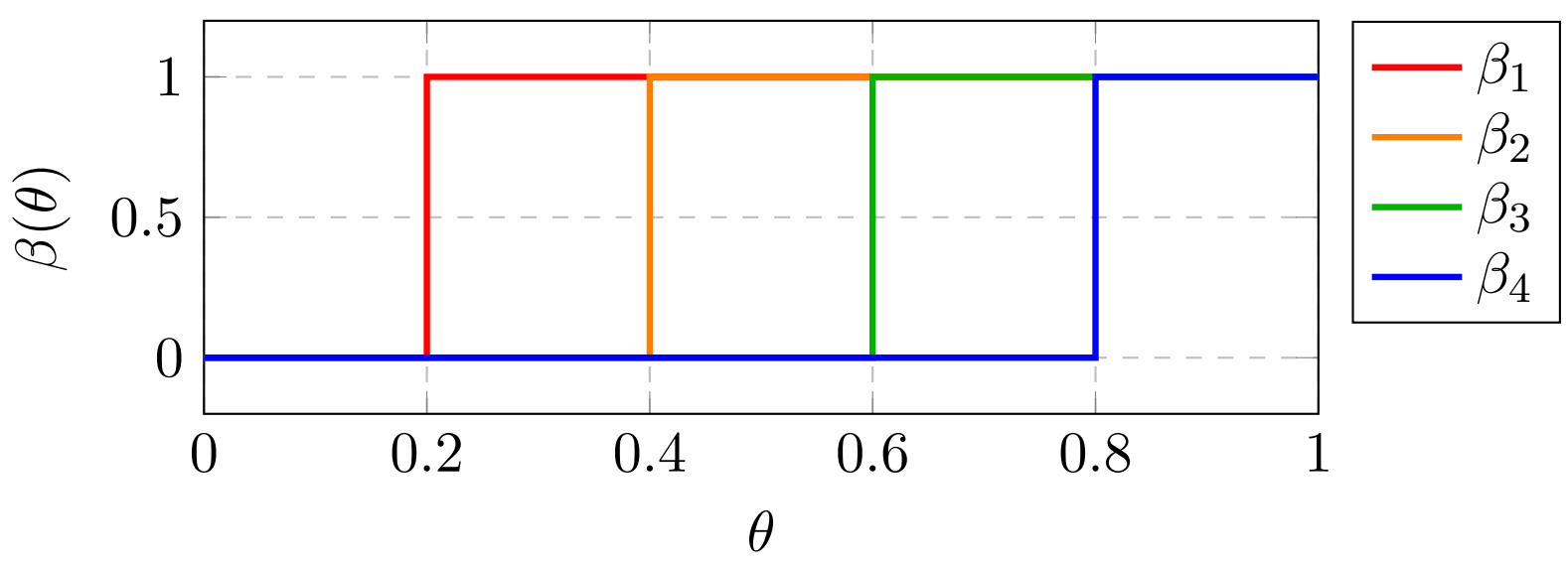

The main idea in the proof of \cref{thm:multiple} is to work with the aggregated outcome of the $k$ ratings, that is, with the aggregation rule of \cref{s:model}. 
For an item of quality $\theta$, let $n(\theta)\in \bc{0,1,\dots,k}$ denote the total number of positive ratings it receives, and let $p(\theta)=(p_0(\theta),\lds,p_k(\theta))$ be its distribution in the simplex $\Delta(\bc{0,1,\dots,k})$, where $p_i(\theta)$ is the probability that the item receives exactly $i$ positive ratings. The platform objective depends on the scales only through the map $\theta\mapsto p(\theta)$, and the proof proceeds in four steps.
\emph{Step 1: relaxation.} Because the scales are nondecreasing, the distribution of $n(\theta)$ is stochastically nondecreasing in $\theta$. We relax the problem to arbitrary families of score distributions with this property and represent each family through its quantile functions.
\emph{Step 2: deterministic and monotone assignments.} A coupling argument shows that such a family never does better than one of the deterministic nondecreasing assignments $\theta\mapsto n(\theta)$ that it is a mixture of, so the problem reduces to partitioning $[0,1]$ into $k+1$ ordered quality intervals.
\emph{Step 3: interval lengths.} For Kendall's $\tau$ function, a Lagrange multiplier computation shows that the intervals should have equal length. This is the only step that depends on the specific weight function.
\emph{Step 4: implementation.} Any such partition is realized by 0-1 step functions with thresholds at the interval boundaries, so the relaxation is tight. The full proof is in \cref{a:multiple}, and Steps 3 and 4 are reused for the adaptive setting in \cref{s:adaptive}.

\paragraph{Non-uniform weight functions.}

Since only Step 3 depends on the specific weight function, for any weight function $w(\theta_1, \theta_2)$ the optimal $k$ scales are still a collection of 0-1 step functions, but how the $k$ thresholds are spaced in $[0,1]$ depends on $w(\theta_1, \theta_2)$: the thresholds minimize the total weight of the pairs of qualities that fall in the same interval, as shown in \cref{eq:general-w} of \cref{a:multiple}. \Cref{a:adaptive-spacing} works out the thresholds for Spearman's $\rho$ function, for which they turn out to be evenly spaced as well, and for a weight function that only values the top half of the items, for which they are not.
Thus, the optimal multiple-scale binary rating system partitions items into $k+1$ ordered quality intervals, so that items in different intervals receive different numbers of positive ratings. We can distinguish items that lie in different intervals with probability 1. Items that lie in the same interval cannot be distinguished because they accumulate the same number of positive ratings.

\paragraph{Optimal objective value for $k$ scales.}

By \cref{thm:multiple}, for Kendall's $\tau$ function the maximum platform objective is achieved using 0-1 step functions with uniformly spaced thresholds in $[0,1]$, so
\begin{align*}
    W^\te{multiple}_k &= 2\int_{\theta_1 < \theta_2} \one\bb{\theta_1, \theta_2 \text{ lie in different intervals}}\,d\theta_1 d\theta_2
    % \\&= 1 - \int_{\theta_1 < \theta_2} \one\bb{\theta_1, \theta_2 \text{ in same interval}}\,d\theta_1 d\theta_2
    = 1 - \fr{1}{k+1}
\end{align*}
since the probability that two items lie in the same quality interval is $\fr{1}{k+1}$. 

% Hence we have shown the following:
    
% \begin{proposition}[Platform objective for multiple scales] \label{thm:multiple-value}
%     The maximum platform objective for $k$ ratings using $k$ scales is $W_k = 1 - \fr{1}{k+1}$.
% \end{proposition}

\paragraph{Comparison with single scale.}

\begin{figure}[!t]
\centering
\begin{tikzpicture}
\begin{axis}[
    width=10.5cm,
    height=4.3cm,
    title={\textbf{Platform objective increases with multiple scales}},
    xlabel={Number of ratings $k$},
    ylabel={Platform objective},
    xmin=0.5, xmax=10.5,
    ymin=0.45, ymax=0.95,
    xtick={1,2,3,4,5,6,7,8,9,10},
    ytick={0.5,0.6,0.7,0.8,0.9},
    legend pos=south east,
    grid=major,
    legend cell align={left}
]

\addplot[
    color=blue,
    mark=*,
    very thick,
]
coordinates {
    (1, 0.5)
    (2, 0.6666667)
    (3, 0.7500000)
    (4, 0.8000000)
    (5, 0.8333333)
    (6, 0.8571429)
    (7, 0.8750000)
    (8, 0.8888889)
    (9, 0.9000000)
    (10, 0.9090909)
};
\addlegendentry{$W^\te{multiple}_k$}

\addplot[
    dotted,
    color=orange,
    mark=*,
    very thick
]
coordinates {
    (1, 0.5)
    (2, 0.6137056)
    (3, 0.6822338)
    (4, 0.7289353)
    (5, 0.7631177)
    (6, 0.7893533)
    (7, 0.8101893)
    (8, 0.8271715)
    (9, 0.8412982)
    (10, 0.8532456)
};
\addlegendentry{$\bar{W}^\te{single}_k$}

\addplot[
    dashed,
    color=purple,
    mark=*,
    very thick
]
coordinates {
    (1, 0.500)
    (2, 0.576)
    (3, 0.637)
    (4, 0.678)
    (5, 0.707)
    (6, 0.730)
    (7, 0.748)
    (8, 0.762)
    (9, 0.775)
    (10, 0.785)
};
\addlegendentry{$\approx W^\te{single}_k$}

\end{axis}
\end{tikzpicture}

\caption{For $k\ge 2$, using a different scale for each rating has a concrete advantage over using the same scale. The approximation for $W^\te{single}_k$ is calculated based on the best discretized scale computed in \Cref{fig:single-scale-step}.}

\label{fig:gap}
\end{figure}
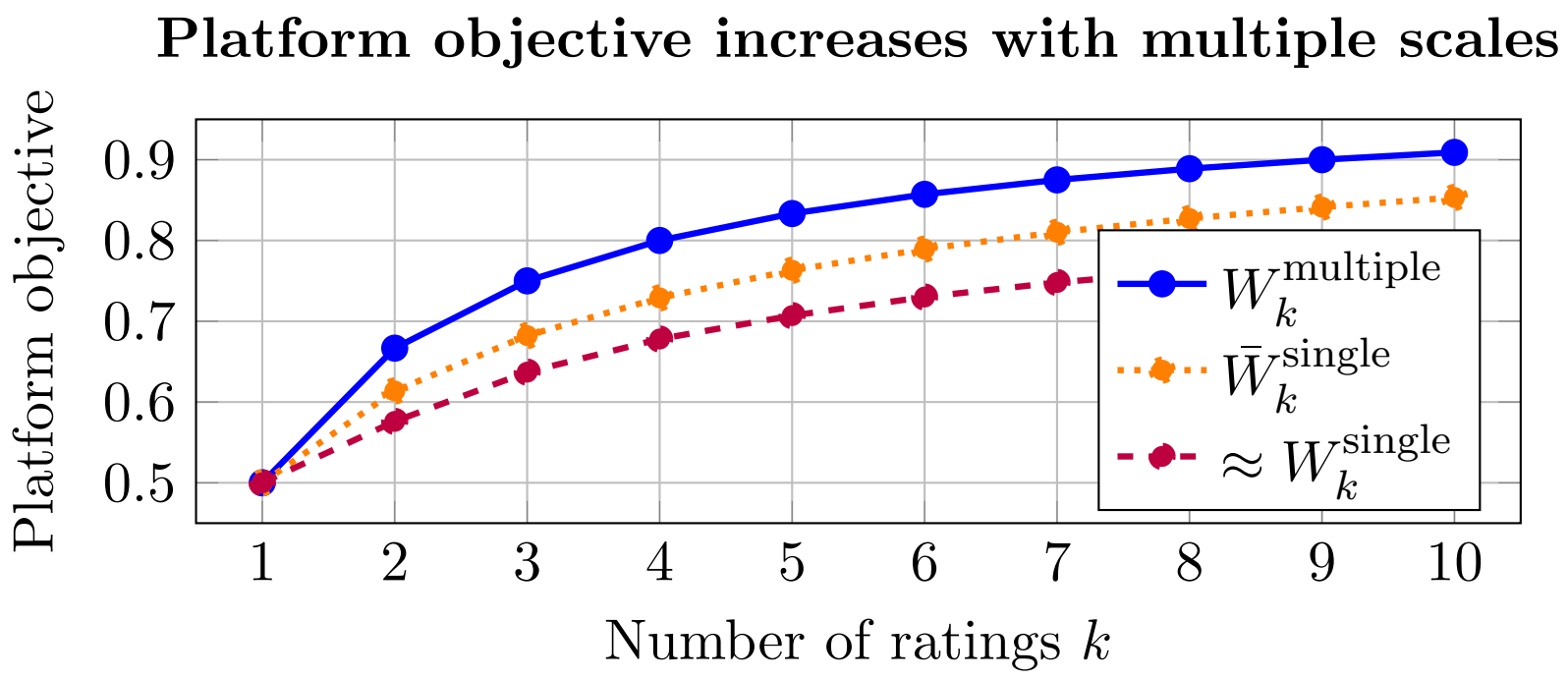

\Cref{fig:gap} compares how much larger the maximum platform objective $W^\te{multiple}_k$ is for multiple scales compared to the maximum objective $W^\te{single}_k$ for a single scale. Recall from \cref{s:model} that computing $W^\te{single}_k$ exactly is intractable, so we employed two estimation methods: the first is to upper bound the platform objective by $\bar{W}^\te{single}_k$ and the second is to estimate $\approx W^\te{single}_k$ by solving for the best discretized scale. For example, at $k=10$ the best discretized single scale achieves $\approx 0.785$, the upper bound is $\bar{W}^\te{single}_{10}\approx 0.853$, and ten specialized scales achieve $W^\te{multiple}_{10} = \fr{10}{11}\approx 0.909$. Because the multiple-scale value exceeds even the upper bound, the gap between the solid and the dotted curves in \cref{fig:gap} is a rigorous lower bound on the advantage of heterogeneous scales, and the gap to the dashed curve is our best estimate of it.

\subsection{Multiple ratings per scale}

\begin{figure}[!t]
\centering
\begin{tikzpicture}
\begin{axis}[
    width=0.335\linewidth,
    height=4.3cm,
    xlabel={$\theta$},
    ylabel={$\beta(\theta)$},
    xlabel shift=-4pt, ylabel shift=-6pt,
    tick label style={font=\scriptsize},
    title style={font=\small},
    legend style={font=\scriptsize, fill opacity=0.8, draw opacity=1, text opacity=1},
    title={\tbf{$\ell=2$ scales}},
    legend pos=north west,
    grid=major
]

\addplot[dashed, color=purple, mark=none, very thick] coordinates {
(0.0000, 0.0000) (0.0204, 0.0000) (0.0408, 0.0000) (0.0612, 0.0000) (0.0816, 0.0000) (0.1020, 0.0140) (0.1224, 0.0300) (0.1429, 0.0450) (0.1633, 0.0600) (0.1837, 0.0770) (0.2041, 0.0950) (0.2245, 0.1150) (0.2449, 0.1360) (0.2653, 0.1590) (0.2857, 0.1840) (0.3061, 0.2100) (0.3265, 0.2370) (0.3469, 0.2650) (0.3673, 0.2950) (0.3878, 0.3250) (0.4082, 0.3560) (0.4286, 0.3870) (0.4490, 0.4190) (0.4694, 0.4510) (0.4898, 0.4840) (0.5102, 0.5160) (0.5306, 0.5490) (0.5510, 0.5810) (0.5714, 0.6130) (0.5918, 0.6440) (0.6122, 0.6750) (0.6327, 0.7050) (0.6531, 0.7350) (0.6735, 0.7630) (0.6939, 0.7900) (0.7143, 0.8160) (0.7347, 0.8410) (0.7551, 0.8640) (0.7755, 0.8850) (0.7959, 0.9050) (0.8163, 0.9230) (0.8367, 0.9400) (0.8571, 0.9550) (0.8776, 0.9700) (0.8980, 0.9860) (0.9184, 1.0000) (0.9388, 1.0000) (0.9592, 1.0000) (0.9796, 1.0000) (1.0000, 1.0000)
};
\addlegendentry{$\beta$}

\addplot[color=green!60!black, mark=none, very thick] coordinates {
(0.0000, 0.0000) (0.0204, 0.0000) (0.0408, 0.0000) (0.0612, 0.0000) (0.0816, 0.0380) (0.1020, 0.0760) (0.1224, 0.1150) (0.1429, 0.1550) (0.1633, 0.2000) (0.1837, 0.2490) (0.2041, 0.3030) (0.2245, 0.3590) (0.2449, 0.4180) (0.2653, 0.4770) (0.2857, 0.5370) (0.3061, 0.5970) (0.3265, 0.6550) (0.3469, 0.7100) (0.3673, 0.7630) (0.3878, 0.8110) (0.4082, 0.8550) (0.4286, 0.8950) (0.4490, 0.9330) (0.4694, 0.9730) (0.4898, 1.0000) (0.5102, 1.0000) (0.5306, 1.0000) (0.5510, 1.0000) (0.5714, 1.0000) (0.5918, 1.0000) (0.6122, 1.0000) (0.6327, 1.0000) (0.6531, 1.0000) (0.6735, 1.0000) (0.6939, 1.0000) (0.7143, 1.0000) (0.7347, 1.0000) (0.7551, 1.0000) (0.7755, 1.0000) (0.7959, 1.0000) (0.8163, 1.0000) (0.8367, 1.0000) (0.8571, 1.0000) (0.8776, 1.0000) (0.8980, 1.0000) (0.9184, 1.0000) (0.9388, 1.0000) (0.9592, 1.0000) (0.9796, 1.0000) (1.0000, 1.0000)
};
\addlegendentry{$\beta_1$}

\addplot[color=blue, mark=none, very thick] coordinates {
(0.0000, 0.0000) (0.0204, 0.0000) (0.0408, 0.0000) (0.0612, 0.0000) (0.0816, 0.0000) (0.1020, 0.0000) (0.1224, 0.0000) (0.1429, 0.0000) (0.1633, 0.0000) (0.1837, 0.0000) (0.2041, 0.0000) (0.2245, 0.0000) (0.2449, 0.0000) (0.2653, 0.0000) (0.2857, 0.0000) (0.3061, 0.0000) (0.3265, 0.0000) (0.3469, 0.0000) (0.3673, 0.0000) (0.3878, 0.0000) (0.4082, 0.0000) (0.4286, 0.0000) (0.4490, 0.0000) (0.4694, 0.0000) (0.4898, 0.0000) (0.5102, 0.0000) (0.5306, 0.0270) (0.5510, 0.0670) (0.5714, 0.1050) (0.5918, 0.1450) (0.6122, 0.1890) (0.6327, 0.2370) (0.6531, 0.2900) (0.6735, 0.3450) (0.6939, 0.4030) (0.7143, 0.4630) (0.7347, 0.5230) (0.7551, 0.5820) (0.7755, 0.6410) (0.7959, 0.6970) (0.8163, 0.7510) (0.8367, 0.8000) (0.8571, 0.8450) (0.8776, 0.8850) (0.8980, 0.9240) (0.9184, 0.9620) (0.9388, 1.0000) (0.9592, 1.0000) (0.9796, 1.0000) (1.0000, 1.0000)
};
\addlegendentry{$\beta_2$}

\end{axis}
\end{tikzpicture}\hspace{-2pt}
\begin{tikzpicture}
\begin{axis}[
    width=0.335\linewidth,
    height=4.3cm,
    xlabel={$\theta$},
    ylabel={$\beta(\theta)$},
    xlabel shift=-4pt, ylabel shift=-6pt,
    tick label style={font=\scriptsize},
    title style={font=\small},
    legend style={font=\scriptsize, fill opacity=0.8, draw opacity=1, text opacity=1},
    title={\tbf{$\ell=3$ scales}},
    legend pos=north west,
    grid=major
]

\addplot[dashed, color=purple, mark=none, very thick] coordinates {
(0.0000, 0.0000) (0.0204, 0.0000) (0.0408, 0.0000) (0.0612, 0.0000) (0.0816, 0.0000) (0.1020, 0.0140) (0.1224, 0.0300) (0.1429, 0.0450) (0.1633, 0.0600) (0.1837, 0.0770) (0.2041, 0.0950) (0.2245, 0.1150) (0.2449, 0.1360) (0.2653, 0.1590) (0.2857, 0.1840) (0.3061, 0.2100) (0.3265, 0.2370) (0.3469, 0.2650) (0.3673, 0.2950) (0.3878, 0.3250) (0.4082, 0.3560) (0.4286, 0.3870) (0.4490, 0.4190) (0.4694, 0.4510) (0.4898, 0.4840) (0.5102, 0.5160) (0.5306, 0.5490) (0.5510, 0.5810) (0.5714, 0.6130) (0.5918, 0.6440) (0.6122, 0.6750) (0.6327, 0.7050) (0.6531, 0.7350) (0.6735, 0.7630) (0.6939, 0.7900) (0.7143, 0.8160) (0.7347, 0.8410) (0.7551, 0.8640) (0.7755, 0.8850) (0.7959, 0.9050) (0.8163, 0.9230) (0.8367, 0.9400) (0.8571, 0.9550) (0.8776, 0.9700) (0.8980, 0.9860) (0.9184, 1.0000) (0.9388, 1.0000) (0.9592, 1.0000) (0.9796, 1.0000) (1.0000, 1.0000)
};
\addlegendentry{$\beta$}

\addplot[color=orange, mark=none, very thick] coordinates {
(0.0000, 0.0000) (0.0204, 0.0000) (0.0408, 0.0000) (0.0612, 0.0340) (0.0816, 0.1030) (0.1020, 0.1680) (0.1224, 0.2370) (0.1429, 0.3110) (0.1633, 0.3920) (0.1837, 0.4770) (0.2041, 0.5640) (0.2245, 0.6470) (0.2449, 0.7250) (0.2653, 0.7960) (0.2857, 0.8620) (0.3061, 0.9260) (0.3265, 1.0000) (0.3469, 1.0000) (0.3673, 1.0000) (0.3878, 1.0000) (0.4082, 1.0000) (0.4286, 1.0000) (0.4490, 1.0000) (0.4694, 1.0000) (0.4898, 1.0000) (0.5102, 1.0000) (0.5306, 1.0000) (0.5510, 1.0000) (0.5714, 1.0000) (0.5918, 1.0000) (0.6122, 1.0000) (0.6327, 1.0000) (0.6531, 1.0000) (0.6735, 1.0000) (0.6939, 1.0000) (0.7143, 1.0000) (0.7347, 1.0000) (0.7551, 1.0000) (0.7755, 1.0000) (0.7959, 1.0000) (0.8163, 1.0000) (0.8367, 1.0000) (0.8571, 1.0000) (0.8776, 1.0000) (0.8980, 1.0000) (0.9184, 1.0000) (0.9388, 1.0000) (0.9592, 1.0000) (0.9796, 1.0000) (1.0000, 1.0000)
};
\addlegendentry{$\beta_1$}

\addplot[color=green!60!black, mark=none, very thick] coordinates {
(0.0000, 0.0000) (0.0204, 0.0000) (0.0408, 0.0000) (0.0612, 0.0000) (0.0816, 0.0000) (0.1020, 0.0000) (0.1224, 0.0000) (0.1429, 0.0000) (0.1633, 0.0000) (0.1837, 0.0000) (0.2041, 0.0000) (0.2245, 0.0000) (0.2449, 0.0000) (0.2653, 0.0000) (0.2857, 0.0000) (0.3061, 0.0000) (0.3265, 0.0000) (0.3469, 0.0000) (0.3673, 0.0340) (0.3878, 0.1010) (0.4082, 0.1630) (0.4286, 0.2280) (0.4490, 0.2990) (0.4694, 0.3770) (0.4898, 0.4580) (0.5102, 0.5420) (0.5306, 0.6230) (0.5510, 0.7010) (0.5714, 0.7720) (0.5918, 0.8370) (0.6122, 0.8990) (0.6327, 0.9660) (0.6531, 1.0000) (0.6735, 1.0000) (0.6939, 1.0000) (0.7143, 1.0000) (0.7347, 1.0000) (0.7551, 1.0000) (0.7755, 1.0000) (0.7959, 1.0000) (0.8163, 1.0000) (0.8367, 1.0000) (0.8571, 1.0000) (0.8776, 1.0000) (0.8980, 1.0000) (0.9184, 1.0000) (0.9388, 1.0000) (0.9592, 1.0000) (0.9796, 1.0000) (1.0000, 1.0000)
};
\addlegendentry{$\beta_2$}

\addplot[color=blue, mark=none, very thick] coordinates {
(0.0000, 0.0000) (0.0204, 0.0000) (0.0408, 0.0000) (0.0612, 0.0000) (0.0816, 0.0000) (0.1020, 0.0000) (0.1224, 0.0000) (0.1429, 0.0000) (0.1633, 0.0000) (0.1837, 0.0000) (0.2041, 0.0000) (0.2245, 0.0000) (0.2449, 0.0000) (0.2653, 0.0000) (0.2857, 0.0000) (0.3061, 0.0000) (0.3265, 0.0000) (0.3469, 0.0000) (0.3673, 0.0000) (0.3878, 0.0000) (0.4082, 0.0000) (0.4286, 0.0000) (0.4490, 0.0000) (0.4694, 0.0000) (0.4898, 0.0000) (0.5102, 0.0000) (0.5306, 0.0000) (0.5510, 0.0000) (0.5714, 0.0000) (0.5918, 0.0000) (0.6122, 0.0000) (0.6327, 0.0000) (0.6531, 0.0000) (0.6735, 0.0000) (0.6939, 0.0740) (0.7143, 0.1380) (0.7347, 0.2040) (0.7551, 0.2750) (0.7755, 0.3530) (0.7959, 0.4360) (0.8163, 0.5230) (0.8367, 0.6080) (0.8571, 0.6890) (0.8776, 0.7630) (0.8980, 0.8320) (0.9184, 0.8970) (0.9388, 0.9660) (0.9592, 1.0000) (0.9796, 1.0000) (1.0000, 1.0000)
};
\addlegendentry{$\beta_3$}

\end{axis}
\end{tikzpicture}\hspace{-2pt}
\begin{tikzpicture}
\begin{axis}[
    width=0.335\linewidth,
    height=4.3cm,
    xlabel={$\theta$},
    ylabel={$\beta(\theta)$},
    xlabel shift=-4pt, ylabel shift=-6pt,
    tick label style={font=\scriptsize},
    title style={font=\small},
    legend style={font=\scriptsize, fill opacity=0.8, draw opacity=1, text opacity=1},
    title={\tbf{$\ell=4$ scales}},
    legend pos=north west,
    grid=major
]

\addplot[dashed, color=purple, mark=none, very thick] coordinates {
(0.0000, 0.0000) (0.0204, 0.0000) (0.0408, 0.0000) (0.0612, 0.0000) (0.0816, 0.0000) (0.1020, 0.0140) (0.1224, 0.0300) (0.1429, 0.0450) (0.1633, 0.0600) (0.1837, 0.0770) (0.2041, 0.0950) (0.2245, 0.1150) (0.2449, 0.1360) (0.2653, 0.1590) (0.2857, 0.1840) (0.3061, 0.2100) (0.3265, 0.2370) (0.3469, 0.2650) (0.3673, 0.2950) (0.3878, 0.3250) (0.4082, 0.3560) (0.4286, 0.3870) (0.4490, 0.4190) (0.4694, 0.4510) (0.4898, 0.4840) (0.5102, 0.5160) (0.5306, 0.5490) (0.5510, 0.5810) (0.5714, 0.6130) (0.5918, 0.6440) (0.6122, 0.6750) (0.6327, 0.7050) (0.6531, 0.7350) (0.6735, 0.7630) (0.6939, 0.7900) (0.7143, 0.8160) (0.7347, 0.8410) (0.7551, 0.8640) (0.7755, 0.8850) (0.7959, 0.9050) (0.8163, 0.9230) (0.8367, 0.9400) (0.8571, 0.9550) (0.8776, 0.9700) (0.8980, 0.9860) (0.9184, 1.0000) (0.9388, 1.0000) (0.9592, 1.0000) (0.9796, 1.0000) (1.0000, 1.0000)
};
\addlegendentry{$\beta$}

\addplot[color=red, mark=none, very thick] coordinates {
(0.0000, 0.0000) (0.0204, 0.0000) (0.0408, 0.0000) (0.0612, 0.0710) (0.0816, 0.1650) (0.1020, 0.2550) (0.1224, 0.3510) (0.1429, 0.4530) (0.1633, 0.5590) (0.1837, 0.6600) (0.2041, 0.7550) (0.2245, 0.8460) (0.2449, 0.9410) (0.2653, 1.0000) (0.2857, 1.0000) (0.3061, 1.0000) (0.3265, 1.0000) (0.3469, 1.0000) (0.3673, 1.0000) (0.3878, 1.0000) (0.4082, 1.0000) (0.4286, 1.0000) (0.4490, 1.0000) (0.4694, 1.0000) (0.4898, 1.0000) (0.5102, 1.0000) (0.5306, 1.0000) (0.5510, 1.0000) (0.5714, 1.0000) (0.5918, 1.0000) (0.6122, 1.0000) (0.6327, 1.0000) (0.6531, 1.0000) (0.6735, 1.0000) (0.6939, 1.0000) (0.7143, 1.0000) (0.7347, 1.0000) (0.7551, 1.0000) (0.7755, 1.0000) (0.7959, 1.0000) (0.8163, 1.0000) (0.8367, 1.0000) (0.8571, 1.0000) (0.8776, 1.0000) (0.8980, 1.0000) (0.9184, 1.0000) (0.9388, 1.0000) (0.9592, 1.0000) (0.9796, 1.0000) (1.0000, 1.0000)
};
\addlegendentry{$\beta_1$}

\addplot[color=orange, mark=none, very thick] coordinates {
(0.0000, 0.0000) (0.0204, 0.0000) (0.0408, 0.0000) (0.0612, 0.0000) (0.0816, 0.0000) (0.1020, 0.0000) (0.1224, 0.0000) (0.1429, 0.0000) (0.1633, 0.0000) (0.1837, 0.0000) (0.2041, 0.0000) (0.2245, 0.0000) (0.2449, 0.0000) (0.2653, 0.0000) (0.2857, 0.0060) (0.3061, 0.1140) (0.3265, 0.2060) (0.3469, 0.3010) (0.3673, 0.4040) (0.3878, 0.5130) (0.4082, 0.6210) (0.4286, 0.7210) (0.4490, 0.8160) (0.4694, 0.9100) (0.4898, 1.0000) (0.5102, 1.0000) (0.5306, 1.0000) (0.5510, 1.0000) (0.5714, 1.0000) (0.5918, 1.0000) (0.6122, 1.0000) (0.6327, 1.0000) (0.6531, 1.0000) (0.6735, 1.0000) (0.6939, 1.0000) (0.7143, 1.0000) (0.7347, 1.0000) (0.7551, 1.0000) (0.7755, 1.0000) (0.7959, 1.0000) (0.8163, 1.0000) (0.8367, 1.0000) (0.8571, 1.0000) (0.8776, 1.0000) (0.8980, 1.0000) (0.9184, 1.0000) (0.9388, 1.0000) (0.9592, 1.0000) (0.9796, 1.0000) (1.0000, 1.0000)
};
\addlegendentry{$\beta_2$}

\addplot[color=green!60!black, mark=none, very thick] coordinates {
(0.0000, 0.0000) (0.0204, 0.0000) (0.0408, 0.0000) (0.0612, 0.0000) (0.0816, 0.0000) (0.1020, 0.0000) (0.1224, 0.0000) (0.1429, 0.0000) (0.1633, 0.0000) (0.1837, 0.0000) (0.2041, 0.0000) (0.2245, 0.0000) (0.2449, 0.0000) (0.2653, 0.0000) (0.2857, 0.0000) (0.3061, 0.0000) (0.3265, 0.0000) (0.3469, 0.0000) (0.3673, 0.0000) (0.3878, 0.0000) (0.4082, 0.0000) (0.4286, 0.0000) (0.4490, 0.0000) (0.4694, 0.0000) (0.4898, 0.0000) (0.5102, 0.0000) (0.5306, 0.0900) (0.5510, 0.1840) (0.5714, 0.2790) (0.5918, 0.3790) (0.6122, 0.4870) (0.6327, 0.5960) (0.6531, 0.6990) (0.6735, 0.7940) (0.6939, 0.8860) (0.7143, 0.9940) (0.7347, 1.0000) (0.7551, 1.0000) (0.7755, 1.0000) (0.7959, 1.0000) (0.8163, 1.0000) (0.8367, 1.0000) (0.8571, 1.0000) (0.8776, 1.0000) (0.8980, 1.0000) (0.9184, 1.0000) (0.9388, 1.0000) (0.9592, 1.0000) (0.9796, 1.0000) (1.0000, 1.0000)
};
\addlegendentry{$\beta_3$}

\addplot[color=blue, mark=none, very thick] coordinates {
(0.0000, 0.0000) (0.0204, 0.0000) (0.0408, 0.0000) (0.0612, 0.0000) (0.0816, 0.0000) (0.1020, 0.0000) (0.1224, 0.0000) (0.1429, 0.0000) (0.1633, 0.0000) (0.1837, 0.0000) (0.2041, 0.0000) (0.2245, 0.0000) (0.2449, 0.0000) (0.2653, 0.0000) (0.2857, 0.0000) (0.3061, 0.0000) (0.3265, 0.0000) (0.3469, 0.0000) (0.3673, 0.0000) (0.3878, 0.0000) (0.4082, 0.0000) (0.4286, 0.0000) (0.4490, 0.0000) (0.4694, 0.0000) (0.4898, 0.0000) (0.5102, 0.0000) (0.5306, 0.0000) (0.5510, 0.0000) (0.5714, 0.0000) (0.5918, 0.0000) (0.6122, 0.0000) (0.6327, 0.0000) (0.6531, 0.0000) (0.6735, 0.0000) (0.6939, 0.0000) (0.7143, 0.0000) (0.7347, 0.0000) (0.7551, 0.0590) (0.7755, 0.1540) (0.7959, 0.2450) (0.8163, 0.3400) (0.8367, 0.4410) (0.8571, 0.5470) (0.8776, 0.6490) (0.8980, 0.7450) (0.9184, 0.8350) (0.9388, 0.9290) (0.9592, 1.0000) (0.9796, 1.0000) (1.0000, 1.0000)
};
\addlegendentry{$\beta_4$}

\end{axis}
\end{tikzpicture}

\caption{The best discretized sets of $\ell$ scales for $\ell\in \bc{2,3,4}$ are graphed alongside the best discretized single scale $\beta$. All graphs use Kendall's $\tau$ function, $k=24$ ratings, and a discretization of $n=50$ steps. For better visualization, we connect the dots between steps instead of graphing the discontinuous step functions themselves.}

\label{fig:multiple-scale-step}
\end{figure}
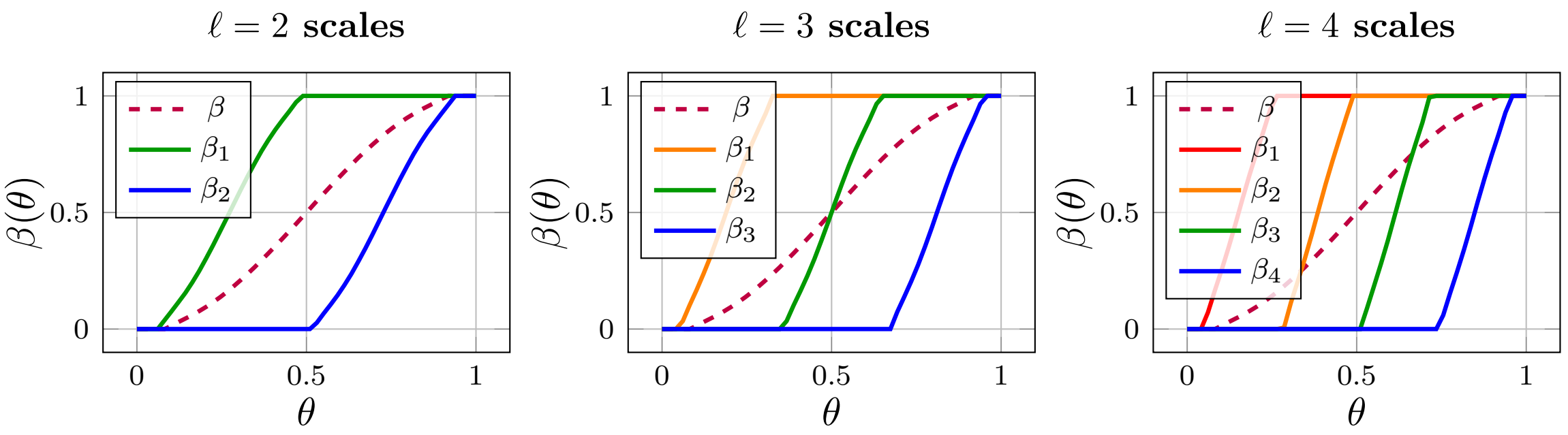

In this section we consider a setting in which we allow multiple rating scales, but not so many as to have one rating per scale. If solving for the best single scale for $k$ ratings is hard, then computing the optimal set of $\ell$ scales each used for $\fr{k}{\ell}$ ratings must be challenging as well. To estimate the optimal configuration of $\ell$ scales, we again employ discretization. We split $[0,1]$ into $n$ equally spaced intervals and solve for the step functions $\beta_1,\lds,\beta_\ell$ that are constant within each interval and that maximize $W_k$. Examples of this numerical computation for $\ell \in \bc{2, 3, 4}$ are shown in \Cref{fig:multiple-scale-step}. 

We make the empirical observation that
the optimal configuration of multiple scales exhibits \emph{specialization}: each scale focuses on distinguishing items from a certain region. Furthermore, as the number $\ell$ of allowed scales increases to $k$, the rating scales $\beta_1,\lds,\beta_\ell$ converge to 0-1 step functions within their respective specialized regions. In the limiting case $\ell = k$, \cref{thm:multiple} tells us that each $\beta_i$ is fully specialized: the job of $\beta_i$ is to exactly distinguish whether an item has quality higher or lower than the $i$-th threshold.

\section{Designing adaptive rating scales} \label{s:adaptive}

In this section, we consider an adaptive setting where $k$ users arrive sequentially, and the scale shown to each user may depend on the binary ratings of all previous users. Our main result is that adaptivity provides an exponential gain in the maximum platform objective $W_k$.

\paragraph{Adaptive model.} The scale $\beta_j$ shown to the $j$-th user may depend on the ratings $(r_1,\lds,r_{j-1})\in\{0,1\}^{j-1}$ of the first $j-1$ users. Equivalently, the design is a complete binary tree of depth $k$: a node at level $j$ is labeled by the nondecreasing scale used for the $j$-th rating, and its two children correspond to the two possible binary responses, which determine the scale used for the next rating. Under scale $\beta_j$, an item of quality $\theta$ receives a positive rating independently with probability $\beta_j(\theta)$, exactly as before. After $k$ ratings, an item's \emph{outcome} is its full response sequence $\omega = (r_1,\lds,r_k)\in \Omega := \{0,1\}^k$, that is, the leaf of the tree that the item reaches, and the platform ranks items by their outcomes according to a total order $\prec$ on $\Omega$ of its choice. The pairwise ranking accuracy between qualities $\theta_1<\theta_2$ is thus
\[
    P_k(\theta_1, \theta_2) := \P{\omega(\theta_1) \prec \omega(\theta_2)} - \P{\omega(\theta_1) \succ \omega(\theta_2)},
\]
and $W_k$ is obtained from $P_k$ exactly as in \cref{s:model}. With $k=2$, for example, the sequences $01$ and $10$ tie under the count-based rule of \cref{s:model}, but they lead to different leaves of the tree and may be ranked differently here. The adaptive model nevertheless contains the non-adaptive one, since a tree whose scales ignore previous responses, with outcomes ordered by their number of ones and ties allowed, recovers the objective of \cref{s:model} exactly.

The adaptive model therefore changes both which questions are asked and how answers are aggregated: items are ranked by their sequence of responses rather than by their number of positive ratings. The next proposition shows that the second change is necessary, so the exponential gain in \cref{thm:adaptive} below should be attributed to the combination of adaptive questions and ranking by the full response sequence, not to adaptivity alone. Fully sequential adaptation also requires each rating to be collected after the previous ones. When ratings arrive in parallel, the batch-adaptive objective at the end of this section quantifies the resulting loss.

\begin{proposition}[Adaptivity with count-based ranking] \label{prop:count-adaptive}
    Let $w$ be Kendall's $\tau$ function. Suppose the platform elicits ratings adaptively with any tree of scales of depth $k$, but still ranks items by the number of positive ratings $n_k(\theta)\in\{0,1,\dots,k\}$. Then $W_k\le 1-\fr{1}{k+1} = W^\te{multiple}_k$, which is already attained by the non-adaptive scales of \cref{thm:multiple}.
\end{proposition}

\begin{proof}
    Any tree of scales induces, for each quality $\theta$, a random variable $n_k(\theta)\in\{0,1,\dots,k\}$. \Cref{lem:relaxation} in \cref{a:multiple} bounds the objective of \emph{every} such collection of random variables, however it is generated, by the objective of the best partition of $[0,1]$ into $k+1$ ordered intervals, which for Kendall's $\tau$ function equals $1-\fr{1}{k+1}$ by \cref{thm:multiple}.
\end{proof}

In words, with only $k+1$ possible scores the platform can distinguish at most $k+1$ groups of items, whatever questions it asks. Adaptivity creates $2^k$ distinguishable outcomes, and the platform must rank by them to benefit. By \cref{lem:relaxation}, the order that the platform uses on the $2^k$ outcomes is immaterial for the optimal value.

\begin{theorem} \label{thm:adaptive}
    Consider any weight function $w(\theta_1, \theta_2)$ that is nonincreasing in $\theta_1$ and nondecreasing in $\theta_2$, such as Kendall's $\tau$ function or Spearman's $\rho$ function. Given $k$ scales $\beta_1,\lds,\beta_k$ to be adaptively chosen for $k$ ratings, where the $j$-th rating uses scale $\beta_j$ and can depend on the responses to the first $j-1$ scales, and given that items are ranked by their response sequence $\omega\in\{0,1\}^k$, the maximum platform objective is achieved by partitioning $[0,1]$ into $2^k$ ordered quality intervals and performing a binary search to determine which of these intervals the item lies in. Concretely, every scale in the tree is a 0-1 step function whose threshold is the middle boundary of the set of intervals still consistent with the responses so far, and the outcomes are ordered lexicographically: $\omega_1\prec\omega_2$ if and only if $\omega_1 < \omega_2$ when read as binary numbers. When $w$ is Kendall's $\tau$ function, the $2^k$ intervals have equal length.
\end{theorem}

\begin{proof}[Proof sketch]
    The full proof is in \cref{a:adaptive} and follows the four steps of the proof of \cref{thm:multiple}. Any tree of scales maps each quality $\theta$ to a distribution $p(\theta)\in\Delta(\Omega)$ over outcomes, so we relax the problem and optimize over arbitrary maps $\theta\mapsto p(\theta)\in\Delta(\Omega)$ and arbitrary orders $\prec$ on $\Omega$. Unlike in the non-adaptive setting, this family need not be monotone in $\theta$ with respect to the order chosen by the platform, so Step 2 of \cref{thm:multiple} is replaced by \cref{lem:relaxation} in \cref{a:multiple}, which shows, for the weight functions in the statement, that the relaxed optimum is still achieved by a deterministic assignment that partitions $[0,1]$ into $2^k$ ordered intervals, one per outcome, and that the choice of order is immaterial. The interval lengths are then chosen as in Step 3 of \cref{thm:multiple}. What is new is the implementation: any partition into $2^k$ ordered intervals is realized by a depth-$k$ tree of 0-1 step functions that binary searches for the interval containing $\theta$, with outcomes ordered lexicographically, so the relaxation is tight.
\end{proof}

\paragraph{Platform objective under Kendall's $\tau$ function.} For Kendall's $\tau$ function, the same Lagrange multipliers analysis as in \cref{thm:multiple} yields that the optimal groups in \cref{thm:adaptive} are a partition of $[0,1]$ into $2^k$ equal length intervals. The binary search then uses threshold $\fr12$ for the first rating, threshold $\fr14$ or $\fr34$ for the second rating depending on the first response, and so on, so that the outcome $\omega(\theta)$ is simply the first $k$ binary digits of $\theta$. Since the probability that two items lie in the same quality interval is $\fr{1}{2^k}$, the maximum platform objective under Kendall's $\tau$ function for $k$ ratings using $k$ adaptive scales is $W^\te{adaptive}_k = 1 - \fr{1}{2^k}$. For $k\ge 2$, this is strictly larger than the optimal objective $W^\te{multiple}_k = 1-\fr{1}{k+1}$ under the non-adaptive model. More specifically, $k$ adaptive ratings have the same distinguishing power as $2^k - 1$ non-adaptive ratings.

\paragraph{Partial adaptivity.} Fully sequential adaptation requires each user's scale to be set after all previous ratings of the same item have arrived, which may be infeasible when many ratings arrive in parallel. The two models above are the endpoints of a spectrum. Suppose ratings arrive in $k/b$ batches of size $b$: the $b$ scales within a batch are set simultaneously but may depend on the responses of all previous batches, and items are ranked by the sequence of per-batch positive counts. We show in \cref{a:partial-adaptive} that the maximum objective under Kendall's $\tau$ function is $1-\fr{1}{(b+1)^{k/b}}$, which interpolates between the non-adaptive value $1-\fr{1}{k+1}$ at $b=k$ and the fully adaptive value $1-\fr{1}{2^k}$ at $b=1$. Even a few rounds of adaptation recover a large fraction of the exponential gain.

\section{Conclusion and discussion}\label{s:conclusion}

When designing rating scales on online platforms, we demonstrate the benefits of using heterogeneous scales on different users. This allows user ratings to specialize in distinguishing certain subsets of items, resulting in overall better accuracy. Multiple scales are simple to implement in practice, for example by randomizing the questions shown to different users.

\paragraph{Implementing rating scales.} A complementary question is how a platform can build a real rating system that elicits binary responses according to a given scale $\beta$. \citet{garg2019} describe and validate one such design. The platform starts with a pool $\cY$ of candidate binary questions that differ in the strictness of their wording and a representative set $\Theta$ of items whose qualities are known. It first estimates, for each question $y\in\cY$ and each item $\theta\in\Theta$, the probability $\psi(\theta,y)$ of a positive response, and then chooses a distribution $H$ over $\cY$ that minimizes $\sum_{\theta\in\Theta}\big|\beta(\theta)-\sum_{y\in\cY}\psi(\theta,y)H(y)\big|$. Showing each user a question drawn independently from $H$ then implements a scale close to $\beta$. The specialized scales of \cref{thm:multiple} are implemented by the same procedure with a different target $\beta_i$ for each user. A 0-1 step function with threshold $\tau_i$ corresponds to concentrating $H$ on the question whose strictness matches $\tau_i$.

\citet{garg2019} validate this procedure on Amazon Mechanical Turk, where $200$ workers rated the English proficiency and argument coherence of ten short paragraphs adapted from TOEFL writing samples published by the Educational Testing Service. The paragraphs span five expert quality levels $\theta\in\{0.1,0.3,0.5,0.7,0.9\}$, with two paragraphs per level. Each worker answered threshold questions of the form ``Would you characterize the performance of this item as [adjective] or better?'' for adjectives ranging from ``Abysmal'' to ``Phenomenal,'' which form the question pool $\cY$. The fitted mixture tracks the target scale closely except for the lowest-quality paragraphs, where the available questions could not separate items that the experts had ranked differently, so the question pool must be rich where the target scale varies. For a Kendall's $\tau$ objective, the design asks most users whether the item is ``Fair'' or better, while objectives that emphasize top items shift mass to the ``Good'' question.

\paragraph{Limitations.}
We discuss the main idealizations in our model and how our results should be interpreted in light of them.

\emph{Noisy and imperfectly implemented scales.} The optimal scales in \cref{thm:multiple,thm:adaptive} are 0-1 step functions, which assume that users can perfectly determine whether the quality of an item is above or below a given threshold $\tau$. Real questions induce smooth response functions, for example $\beta_i(\theta) = \sigma(a(\theta-\tau_i))$ for a sigmoid $\sigma$ and a finite slope $a$, so our results should be read as design principles rather than literal implementation targets. Two of our findings do not depend on this idealization. First, \cref{thm:splitting} applies to \emph{any} scale, however noisy: splitting a smooth, non-constant scale into two specialized smooth scales strictly improves the objective. Second, step functions are only optimal when the number of scales equals the number of ratings. When $\ell < k$ scales are shared among $k$ ratings, the best discretized scales in \cref{fig:multiple-scale-step} are themselves smooth and specialized, and they continue to outperform the best single scale. What we do not characterize is how fast the gains in \cref{thm:multiple,thm:adaptive} degrade as the realized scales depart from step functions, for instance as the slope $a$ decreases. We leave such a sensitivity analysis to future work.

\emph{User heterogeneity.} In our model, the probability $\beta_i(\theta)$ of a positive rating depends on the item and on the question, but not on the user. In practice, different users respond differently to the same question: one user may consider an item their best experience while another finds it only mildly positive. Our results still apply when users are matched to items at random, because the rating of a random user is then $\Bern{\beta_q(\theta)}$, where $\beta_q(\theta)$ is the response probability of question $q$ averaged over the user population. This population scale is what a platform estimates from data \citep{garg2019}, and no calibration of individual users is required. Averaging over users is also one reason why realized scales are smoother than step functions. What our theory does not capture is the information in \emph{which} user provided a rating: scales that depend on the user, for example through observable covariates, could specialize further, and we leave this extension to future work.

\emph{User experience and fairness.} Heterogeneous scales are implemented by randomizing the questions shown to different users. Since the randomization is independent of the item and of the user, every seller and every user faces the same distribution of questions in expectation, so no item is systematically evaluated under a stricter question than others. A platform may nonetheless wish to limit how different the questions shown to different users can be, but \cref{thm:splitting} shows that even two different questions already yield a strict improvement.

\paragraph{Open questions.}
In the non-adaptive setting we follow prior work and aggregate ratings by their empirical mean, while in the adaptive setting we rank by the full response sequence. In both cases, the aggregation rule is fixed and the scales are optimized against it. It is an open question to derive better aggregation methods, and moreover, to jointly design the rating scales and the aggregation methods.
More generally, in classic machine learning paradigms, the data distributions are usually assumed exogenous and fixed. The idea of changing the distribution under which people provide data applies more broadly. For example, it would be interesting to jointly design an online recommendation algorithm and a rating scale, which may itself be adaptive.

\section*{Acknowledgments}

Alec Sun was supported by an NSF Graduate Research Fellowship.

\newpage

\appendix
\section{Appendix}
\subsection{Discretized solution to the optimal single scale}
    
% The second method to estimate $W_k$ is to compute the best \emph{discretized} single scale. The results of this computation for $n=50$ are shown in \Cref{fig:single-scale-step}.

\begin{figure}[h]
\centering
\begin{tikzpicture}
\begin{axis}[
    width=12cm,
    height=10cm,
    xmin=0, xmax=1,
    xlabel={$\theta$},
    ylabel={$\beta(\theta)$},
    title={\tbf{Best discretized single scale}},
    legend pos=north west,
    grid=major
]

\addplot[color=pink, mark=none] coordinates {
(0.0000, 0.0000) (0.0204, 0.0000) (0.0408, 0.0000) (0.0612, 0.0000) (0.0816, 0.0000) (0.1020, 0.0000) (0.1224, 0.0000) (0.1429, 0.0000) (0.1633, 0.0000) (0.1837, 0.0000) (0.2041, 0.0000) (0.2245, 0.0000) (0.2449, 0.0000) (0.2653, 0.0000) (0.2857, 0.0000) (0.3061, 0.0000) (0.3265, 0.0000) (0.3469, 0.0000) (0.3673, 0.0000) (0.3878, 0.0000) (0.4082, 0.0000) (0.4286, 0.0000) (0.4490, 0.0000) (0.4694, 0.0000) (0.4898, 0.0000) (0.5102, 1.0000) (0.5306, 1.0000) (0.5510, 1.0000) (0.5714, 1.0000) (0.5918, 1.0000) (0.6122, 1.0000) (0.6327, 1.0000) (0.6531, 1.0000) (0.6735, 1.0000) (0.6939, 1.0000) (0.7143, 1.0000) (0.7347, 1.0000) (0.7551, 1.0000) (0.7755, 1.0000) (0.7959, 1.0000) (0.8163, 1.0000) (0.8367, 1.0000) (0.8571, 1.0000) (0.8776, 1.0000) (0.8980, 1.0000) (0.9184, 1.0000) (0.9388, 1.0000) (0.9592, 1.0000) (0.9796, 1.0000) (1.0000, 1.0000)
};
\addlegendentry{$k = 1$ (0.500)}

\addplot[color=red, mark=none] coordinates {
(0.0000, 0.0000) (0.0204, 0.0000) (0.0408, 0.0000) (0.0612, 0.0000) (0.0816, 0.0000) (0.1020, 0.0000) (0.1224, 0.0000) (0.1429, 0.0000) (0.1633, 0.0000) (0.1837, 0.0000) (0.2041, 0.0000) (0.2245, 0.0000) (0.2449, 0.0000) (0.2653, 0.0000) (0.2857, 0.0070) (0.3061, 0.0710) (0.3265, 0.1290) (0.3469, 0.1810) (0.3673, 0.2300) (0.3878, 0.2760) (0.4082, 0.3190) (0.4286, 0.3610) (0.4490, 0.4010) (0.4694, 0.4410) (0.4898, 0.4800) (0.5102, 0.5200) (0.5306, 0.5590) (0.5510, 0.5990) (0.5714, 0.6390) (0.5918, 0.6810) (0.6122, 0.7240) (0.6327, 0.7700) (0.6531, 0.8190) (0.6735, 0.8710) (0.6939, 0.9290) (0.7143, 0.9930) (0.7347, 1.0000) (0.7551, 1.0000) (0.7755, 1.0000) (0.7959, 1.0000) (0.8163, 1.0000) (0.8367, 1.0000) (0.8571, 1.0000) (0.8776, 1.0000) (0.8980, 1.0000) (0.9184, 1.0000) (0.9388, 1.0000) (0.9592, 1.0000) (0.9796, 1.0000) (1.0000, 1.0000)
};
\addlegendentry{$k = 2$ (0.576)}

\addplot[color=orange, mark=none] coordinates {
(0.0000, 0.0000) (0.0204, 0.0000) (0.0408, 0.0000) (0.0612, 0.0000) (0.0816, 0.0000) (0.1020, 0.0000) (0.1224, 0.0000) (0.1429, 0.0000) (0.1633, 0.0000) (0.1837, 0.0000) (0.2041, 0.0000) (0.2245, 0.0270) (0.2449, 0.0740) (0.2653, 0.1160) (0.2857, 0.1540) (0.3061, 0.1900) (0.3265, 0.2240) (0.3469, 0.2580) (0.3673, 0.2910) (0.3878, 0.3240) (0.4082, 0.3560) (0.4286, 0.3880) (0.4490, 0.4200) (0.4694, 0.4520) (0.4898, 0.4840) (0.5102, 0.5160) (0.5306, 0.5480) (0.5510, 0.5800) (0.5714, 0.6120) (0.5918, 0.6440) (0.6122, 0.6760) (0.6327, 0.7090) (0.6531, 0.7420) (0.6735, 0.7760) (0.6939, 0.8100) (0.7143, 0.8460) (0.7347, 0.8840) (0.7551, 0.9260) (0.7755, 0.9730) (0.7959, 1.0000) (0.8163, 1.0000) (0.8367, 1.0000) (0.8571, 1.0000) (0.8776, 1.0000) (0.8980, 1.0000) (0.9184, 1.0000) (0.9388, 1.0000) (0.9592, 1.0000) (0.9796, 1.0000) (1.0000, 1.0000)
};
\addlegendentry{$k = 4$ (0.678)}

\addplot[color=yellow, mark=none] coordinates {
(0.0000, 0.0000) (0.0204, 0.0000) (0.0408, 0.0000) (0.0612, 0.0000) (0.0816, 0.0000) (0.1020, 0.0000) (0.1224, 0.0000) (0.1429, 0.0000) (0.1633, 0.0190) (0.1837, 0.0490) (0.2041, 0.0760) (0.2245, 0.1020) (0.2449, 0.1270) (0.2653, 0.1520) (0.2857, 0.1780) (0.3061, 0.2050) (0.3265, 0.2320) (0.3469, 0.2600) (0.3673, 0.2890) (0.3878, 0.3190) (0.4082, 0.3510) (0.4286, 0.3830) (0.4490, 0.4160) (0.4694, 0.4490) (0.4898, 0.4830) (0.5102, 0.5170) (0.5306, 0.5510) (0.5510, 0.5840) (0.5714, 0.6170) (0.5918, 0.6490) (0.6122, 0.6810) (0.6327, 0.7110) (0.6531, 0.7400) (0.6735, 0.7680) (0.6939, 0.7950) (0.7143, 0.8220) (0.7347, 0.8480) (0.7551, 0.8730) (0.7755, 0.8980) (0.7959, 0.9240) (0.8163, 0.9510) (0.8367, 0.9810) (0.8571, 1.0000) (0.8776, 1.0000) (0.8980, 1.0000) (0.9184, 1.0000) (0.9388, 1.0000) (0.9592, 1.0000) (0.9796, 1.0000) (1.0000, 1.0000)
};
\addlegendentry{$k = 8$ (0.762)}

\addplot[color=green, mark=none] coordinates {
(0.0000, 0.0000) (0.0204, 0.0000) (0.0408, 0.0000) (0.0612, 0.0000) (0.0816, 0.0000) (0.1020, 0.0000) (0.1224, 0.0180) (0.1429, 0.0370) (0.1633, 0.0550) (0.1837, 0.0740) (0.2041, 0.0930) (0.2245, 0.1130) (0.2449, 0.1340) (0.2653, 0.1570) (0.2857, 0.1810) (0.3061, 0.2070) (0.3265, 0.2340) (0.3469, 0.2630) (0.3673, 0.2920) (0.3878, 0.3230) (0.4082, 0.3540) (0.4286, 0.3860) (0.4490, 0.4180) (0.4694, 0.4510) (0.4898, 0.4840) (0.5102, 0.5160) (0.5306, 0.5490) (0.5510, 0.5820) (0.5714, 0.6140) (0.5918, 0.6460) (0.6122, 0.6770) (0.6327, 0.7080) (0.6531, 0.7370) (0.6735, 0.7660) (0.6939, 0.7930) (0.7143, 0.8190) (0.7347, 0.8430) (0.7551, 0.8660) (0.7755, 0.8870) (0.7959, 0.9070) (0.8163, 0.9260) (0.8367, 0.9450) (0.8571, 0.9630) (0.8776, 0.9820) (0.8980, 1.0000) (0.9184, 1.0000) (0.9388, 1.0000) (0.9592, 1.0000) (0.9796, 1.0000) (1.0000, 1.0000)
};
\addlegendentry{$k = 16$ (0.828)}

\addplot[color=blue, mark=none] coordinates {
(0.0000, 0.0000) (0.0204, 0.0000) (0.0408, 0.0000) (0.0612, 0.0000) (0.0816, 0.0070) (0.1020, 0.0210) (0.1224, 0.0340) (0.1429, 0.0470) (0.1633, 0.0620) (0.1837, 0.0780) (0.2041, 0.0960) (0.2245, 0.1160) (0.2449, 0.1380) (0.2653, 0.1610) (0.2857, 0.1860) (0.3061, 0.2120) (0.3265, 0.2390) (0.3469, 0.2670) (0.3673, 0.2960) (0.3878, 0.3260) (0.4082, 0.3570) (0.4286, 0.3880) (0.4490, 0.4200) (0.4694, 0.4520) (0.4898, 0.4840) (0.5102, 0.5160) (0.5306, 0.5480) (0.5510, 0.5800) (0.5714, 0.6120) (0.5918, 0.6430) (0.6122, 0.6740) (0.6327, 0.7040) (0.6531, 0.7330) (0.6735, 0.7610) (0.6939, 0.7880) (0.7143, 0.8140) (0.7347, 0.8390) (0.7551, 0.8620) (0.7755, 0.8840) (0.7959, 0.9040) (0.8163, 0.9220) (0.8367, 0.9380) (0.8571, 0.9530) (0.8776, 0.9660) (0.8980, 0.9790) (0.9184, 0.9930) (0.9388, 1.0000) (0.9592, 1.0000) (0.9796, 1.0000) (1.0000, 1.0000)
};
\addlegendentry{$k = 32$ (0.876)}

\addplot[color=purple, mark=none] coordinates {
(0.0000, 0.0000) (0.0204, 0.0000) (0.0408, 0.0000) (0.0612, 0.0060) (0.0816, 0.0160) (0.1020, 0.0250) (0.1224, 0.0370) (0.1429, 0.0490) (0.1633, 0.0640) (0.1837, 0.0810) (0.2041, 0.1000) (0.2245, 0.1200) (0.2449, 0.1410) (0.2653, 0.1640) (0.2857, 0.1890) (0.3061, 0.2140) (0.3265, 0.2410) (0.3469, 0.2690) (0.3673, 0.2980) (0.3878, 0.3280) (0.4082, 0.3580) (0.4286, 0.3890) (0.4490, 0.4200) (0.4694, 0.4520) (0.4898, 0.4840) (0.5102, 0.5160) (0.5306, 0.5480) (0.5510, 0.5800) (0.5714, 0.6110) (0.5918, 0.6420) (0.6122, 0.6720) (0.6327, 0.7020) (0.6531, 0.7310) (0.6735, 0.7590) (0.6939, 0.7860) (0.7143, 0.8110) (0.7347, 0.8360) (0.7551, 0.8590) (0.7755, 0.8800) (0.7959, 0.9000) (0.8163, 0.9190) (0.8367, 0.9360) (0.8571, 0.9510) (0.8776, 0.9630) (0.8980, 0.9750) (0.9184, 0.9840) (0.9388, 0.9940) (0.9592, 1.0000) (0.9796, 1.0000) (1.0000, 1.0000)
};
\addlegendentry{$k = 64$ (0.911)}

% \addplot[color=pink, mark=none] coordinates {
% (0.025,0.0) (0.075,0.0) (0.125,0.0) (0.175,0.044) (0.225,0.103) (0.275,0.161) (0.325,0.226) (0.375,0.298) (0.425,0.376) (0.475,0.458) (0.525,0.542) (0.575,0.624) (0.625,0.702) (0.675,0.774) (0.725,0.839) (0.775,0.897) (0.825,0.956) (0.875,1.0) (0.925,1.0) (0.975,1.0)
% };
% \addlegendentry{$k = 10$ (0.784)}

\end{axis}
\end{tikzpicture}

\caption{The optimal single scale for Kendall's $\tau$ function discretized to $n=50$ steps is graphed for all numbers $k\in \bc{1, 2, 4, 8, 16, 32, 64}$ of ratings. The legend shows the platform objective of each discretized scale. For better visualization, we connect the dots between steps instead of graphing the discontinuous step functions themselves.}

\label{fig:single-scale-step}

\end{figure}
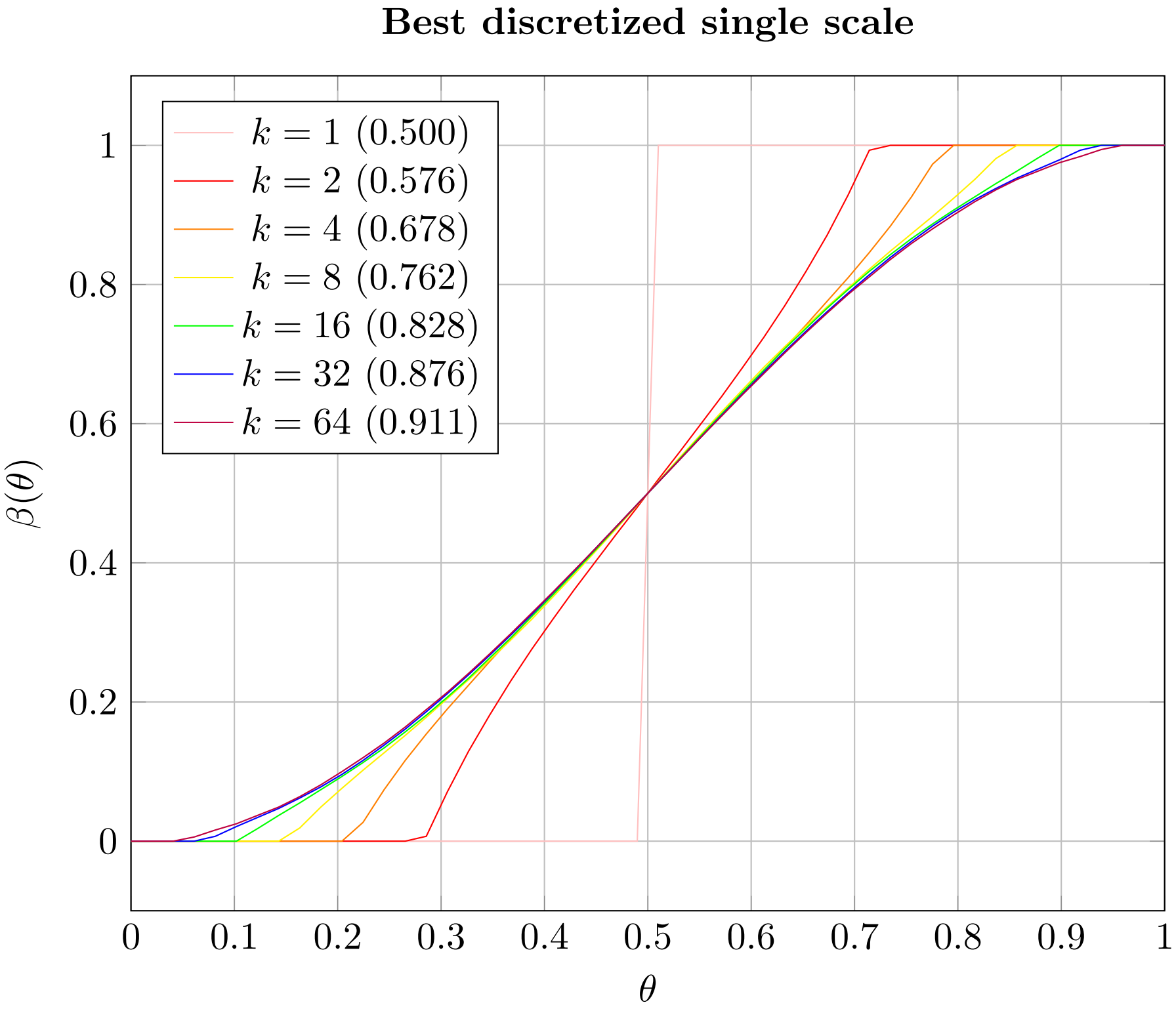

\subsection{Proof of \Cref{thm:single-value}} \label{a:single}

Recall from \cref{s:model} that we have two ways to estimate the platform objective for a single scale. The first is by showing a theoretical upper bound $\bar{W}^\te{single}_k$, and the second is by solving for the best \emph{discretized} single scale.

To prove the upper bound in \cref{thm:single-value}, the idea is to derive a \emph{pairwise} bound for every pair of item qualities. Intuitively, for two quality levels $i<j$ with positive-rating probabilities $p_i\le p_j$, if the difference $\delta=p_j-p_i$ is small, then it is
hard to correctly rank these two levels using repeated noisy ratings.

\begin{lemma} \label{lem:pairwise_ub}
Let $0\le p_i \le p_j \le 1$ and $\delta = p_j - p_i$. Let
\[
S_i:=\sum_{r=1}^k X_{i,r}, \qquad S_j:=\sum_{r=1}^k X_{j,r},
\]
where $X_{i,r} \stackrel{\mathrm{iid}}{\sim} \mathrm{Bernoulli}(p_i)$ and $X_{j,r} \stackrel{\mathrm{iid}}{\sim} \mathrm{Bernoulli}(p_j)$. Then, we have
\[
\mathbb{E}\big[\sgn(S_j-S_i)\big]
\le
1-(1-\delta)^k.
\]
\end{lemma}

\begin{proof}
To bound the expected sign of $S_j - S_i$, we decompose $S_j - S_i$ into two random variables, one with zero mean and one that captures the bias due to the gap $\delta$. 
For each round $r \in [k]$, define
\[
D_r:=X_{j,r}-X_{i,r}\in\{-1,0,1\}.
\]
Then we have $\mathbb{P}(D_r=1)=p_j(1-p_i)$
and $\mathbb{P}(D_r=-1)=p_i(1-p_j)$.
Let $b=p_i(1-p_j)$. Since $\delta = p_j - p_i$, we have 
\[
\mathbb{P}(D_r=1)=\delta+b,\qquad
\mathbb{P}(D_r=-1)=b,\qquad
\mathbb{P}(D_r=0)=1-\delta-2b.
\]

We first decompose the random variable $D_r$ into a bias term and a symmetric term. We define the random variable
$G_r \sim \mathrm{Bernoulli}(\delta)$.
Independently of $G_r$, let $A_r\in\{-1,0,1\}$ be a random variable with the symmetric distribution
\[
\mathbb{P}(A_r=1)=\mathbb{P}(A_r=-1)=\frac{b}{1-\delta},
\qquad
\mathbb{P}(A_r=0)=1-\frac{2b}{1-\delta}.
\]
This distribution is well-defined since $0 \leq 2b/(1-\delta) \leq 1$ for any $0\le p_i \le p_j \le 1$.  
Since $G_r + (1-G_r)A_r$ has the same distribution as $D_r$, we may realize
\[
D_r = G_r + (1-G_r)A_r.
\]

Hence, summing over all rounds, the difference between the two scores is 
\[
S_j-S_i=\sum_{r=1}^k D_r = G+A,
\]
where
\[
G:=\sum_{r=1}^k G_r \sim \mathrm{Binomial}(k,\delta), \qquad A:=\sum_{r=1}^k (1-G_r)A_r.
\]
Note that conditional on $(G_1,\dots,G_k)$, the random variable $A$ is symmetric about $0$, since it is a sum of independent symmetric terms. Therefore, we have
\[
\mathbb{E}\big[\operatorname{sgn}(G+A)\mid G_1,\dots,G_k\big]
=
0
\quad\text{if } G=0,
\]
while trivially
\[
\mathbb{E}\big[\operatorname{sgn}(G+A)\mid G_1,\dots,G_k\big]
\le 1
\quad\text{if } G>0.
\]
Thus, we obtain
\[
\mathbb{E}\big[\operatorname{sgn}(S_j-S_i)\big]
\le
\mathbb{P}(G>0)
=
1-(1-\delta)^k.
\]
\end{proof}

We now use \cref{lem:pairwise_ub} to prove \cref{thm:single-value}.

\begin{proof}[Proof of \cref{thm:single-value}]
Fix any nondecreasing scale $\beta$ and discretize it to $n$ levels $p_i := \beta(i/n)$, so that
\[
0\le p_1\le p_2\le \cdots \le p_n\le 1.
\]
By \cref{lem:pairwise_ub}, for every pair $i<j$, we have $\mathbb{E}\big[\sgn(S_j-S_i)\big]
\le
\phi_k(p_j-p_i)$, where $\phi_k(x):=1-(1-x)^k$.

Therefore, the platform's objective on the discretized scale is 
\[
\frac{1}{\binom n2}
\sum_{1\le i<j\le n}
\mathbb{E}\big[\sgn(S_j-S_i)\big]
\le
\frac{1}{\binom n2}
\sum_{1\le i<j\le n}
\phi_k(p_j-p_i).
\]

We group pairs according to their index gap $\ell=j-i$. Then
\[
\sum_{1\le i<j\le n}\phi_k(p_j-p_i)
=
\sum_{\ell=1}^{n-1}
\sum_{i=1}^{n-\ell}
\phi_k(p_{i+\ell}-p_i).
\]
Since $\phi_k$ is increasing and concave on $[0,1]$, by Jensen's inequality, we have
\[
\sum_{i=1}^{n-\ell}\phi_k(p_{i+\ell}-p_i)
\le
(n-\ell)\,
\phi_k\left(
\frac{1}{n-\ell}
\sum_{i=1}^{n-\ell}(p_{i+\ell}-p_i)
\right).
\]
Moreover, because $0\le p_i\le 1$ for all $i$, when $\ell \le n/2$, so that $n-\ell\ge\ell$, we have 
\[
\sum_{i=1}^{n-\ell}(p_{i+\ell}-p_i)
=
\sum_{i=n-\ell+1}^{n}p_i
-
\sum_{i=1}^{\ell}p_i
\le \ell,
\]
and when $\ell > n/2$, so that $n-\ell<\ell$, we have
\[
\sum_{i=1}^{n-\ell}(p_{i+\ell}-p_i)
=
\sum_{i=\ell+1}^{n}p_i
-
\sum_{i=1}^{n-\ell}p_i
\le n-\ell.
\]
Thus, we have 
\[
\frac{1}{n-\ell}
\sum_{i=1}^{n-\ell}(p_{i+\ell}-p_i)
\le
\frac{\min\{\ell, n-\ell\}}{n-\ell}.
\]
Hence, we obtain
\[
\sum_{i=1}^{n-\ell}\phi_k(p_{i+\ell}-p_i)
\le
(n-\ell)\,
\phi_k\left(
\min\left\{1,\frac{\ell}{n-\ell}\right\}
\right).
\]
Therefore, our objective is at most
\[
U_{n,k}
:=
\frac{1}{\binom n2}
\sum_{\ell=1}^{n-1}
(n-\ell)
\left[
1-
\left(
1-\min\left\{1,\frac{\ell}{n-\ell}\right\}
\right)^k
\right].
\]

Since any scale is approximated arbitrarily well by its discretizations, taking $n\to\infty$ gives the bound for the original scale, where
\[
\lim_{n\to\infty}U_{n,k}
=
2\int_0^1
(1-x)
\left[
1-
\left(
1-\min\left\{1,\frac{x}{1-x}\right\}
\right)^k
\right]dx.
\]
For $x\le 1/2$,
\[
1-\frac{x}{1-x}
=
\frac{1-2x}{1-x},
\]
while for $x\ge 1/2$, the minimum term equals $1$. Hence
\begin{align*}
\lim_{n\to\infty}U_{n,k}
&=
2\int_0^{1/2}
(1-x)
\left[
1-
\left(\frac{1-2x}{1-x}\right)^k
\right]dx
+
2\int_{1/2}^1(1-x)\,dx \\
&=
1
-
2\int_0^{1/2}
\frac{(1-2x)^k}{(1-x)^{k-1}}\,dx.
\end{align*}
Finally, using the change of variables $y=1-2x$, we get
\[
2\int_0^{1/2}
\frac{(1-2x)^k}{(1-x)^{k-1}}\,dx
=
2^{k-1}
\int_0^1
\frac{y^k}{(1+y)^{k-1}}\,dy.
\]
Therefore, our objective is upper bounded by 
\[
\bar{W}^\te{single}_k = 
1-2^{k-1}
\int_0^1
\frac{y^k}{(1+y)^{k-1}}\,dy.
\]
Finally, we note that this bound is strictly smaller than $W^\te{multiple}_k = 1-\frac{1}{k+1}$ for every $k\ge 2$, since
\begin{align*}
    2^{k-1}\int_0^1 \frac{y^k}{(1+y)^{k-1}}\,\dd y 
    &= \int_0^1\left(\frac{2}{1+y}\right)^{k-1}y^k\,\dd y\\
    &> \int_0^1 y^k\,\dd y = \frac{1}{k+1},
\end{align*}
where the inequality uses $2/(1+y)>1$ for $y<1$. This completes the proof.
\end{proof}

\subsection{Proof of \Cref{thm:splitting}} \label{a:splitting}

As discussed in \cref{s:splitting}, to prove \cref{thm:splitting} it suffices to show that $\widetilde K_{2m}(p,q)\ge K_{2m}(p,q)$ for all $p < q$. We do so via a sequence of rating hybrids: we start with the original \(k = 2m\) copies of $\beta$-scales and replace a pair of $\beta$-scales by one $\beta^+$-scale and one $\beta^-$-scale, one pair at a time. We claim that every such replacement weakly increases the objective $W_k$. Define
\[
a:=q(1-p),
\quad
c:=p(1-q),
\quad
r:=\frac{a}{c}=\frac{q(1-p)}{p(1-q)}>1,
\]
\[
U:=\Bern{q} - \Bern{p},
\]
\[
W:=\Bern{B^+(q)} + \Bern{B^-(q)} - \Bern{B^+(p)} - \Bern{B^-(p)},
\]
where all Bernoulli random variables are independent. Here $U$ is the difference between the ratings of the two items under one copy of the original scale $\beta$, and $W$ is the difference between their total ratings under one $\beta^+$-scale and one $\beta^-$-scale. Note that $\E{U} = \E{W} = q-p$: the split preserves the expected number of positive ratings, and the gain comes from the change in the distribution around this mean.

The ratio $r$ is well defined when $c>0$, that is, when $0<p$ and $q<1$. In the boundary cases $p=0$ or $q=1$ we have $c=0$, and then every negative term in $U$ and $W$ vanishes. Hence $U\ge0$ and $W\ge0$ almost surely, so that $\P{U=-t}=\P{W=-t}=0$ for all $t\ge1$. In these cases the ratio bounds below hold trivially with the convention $r=\infty$, every sum below that involves $\P{D=-t}$ for $t\ge1$ vanishes, and the proofs simplify accordingly. We therefore assume $0<p<q<1$ from now on and leave the routine adjustments to the reader.

We introduce three helpful technical lemmas.

\begin{lemma}[One pair bias]
\label{lem:block-ratio}
For every integer \(t\ge1\),
\[
\P{U=t}\ge r^t\cd\P{U=-t},
\qquad
\P{W=t}\ge r^t\cd\P{W=-t}.
\]

\end{lemma}
%Note that the two claims hold trivially for every integer $t\ge 2$ and $t\ge 3$, repsectively.

\begin{proof}
The claim for \(U\) is immediate: \(U\in\{-1,0,1\}\) and
\[
\P{U=1}=q(1-p)=a=r\cd c=r\cd\P{U=-1}.
\]
For \(t\ge2\), both \(\P{U=t}\) and \(\P{U=-t}\) are zero.

For \(W\), we distinguish three cases.

\tbf{Case 1: \(0\le p< q\le \tfrac12\).}
Then \(B^-(p)=B^-(q)=0\),
\(B^+(p)=2p\), and \(B^+(q)=2q\). Hence we have
\[
W=\gamma_q-\gamma_p,
\qquad
\gamma_q\sim\Bern{2q},\ \gamma_p\sim\Bern{2p}.
\]
Hence \(W\in\{-1,0,1\}\) and
\[
\frac{\P{W=1}}{\P{W=-1}}
=
\frac{2q(1-2p)}{2p(1-2q)}
\ge
\frac{q(1-p)}{p(1-q)}
=
r,
\]
where the inequality is from \((1-2p)(1-q)-(1-p)(1-2q)=q-p\ge0 \). If $q=\tfrac12$, then $\P{W=-1}=0$ and there is nothing to prove.
Thus \(\P{W=1}\ge r\cd\P{W=-1}\). Since \(\P{W=\pm t}=0\), the bound also holds
for \(t\geq 2\).

\tbf{Case 2: \(\tfrac12\le p< q\le1\).}
Then \(B^+(p)=B^+(q)=1\) and
\[
B^-(p)=2p-1,
\qquad
B^-(q)=2q-1.
\]
So
\[
W=\eta_q-\eta_p,
\qquad
\eta_q\sim\Bern{2q-1},\ \eta_p\sim\Bern{2p-1}.
\]
Hence \(W\in\{-1,0,1\}\) and
\[
\frac{\P{W=1}}{\P{W=-1}}
=
\frac{(2q-1)(1-p)}{(2p-1)(1-q)}
\ge
\frac{q(1-p)}{p(1-q)}
=
r,
\]
where the inequality is from 
\(p(2q-1)-q(2p-1)=q-p\ge0\). If $p=\tfrac12$, then $\P{W=-1}=0$ and there is nothing to prove.
Again \(\P{W=\pm t}=0\) for all $t \geq 2$, so the bound holds for all \(t\ge1\).

\tbf{Case 3: \(0\le p\le \tfrac12\le q\le1\).}
Then
\[
B^+(p)=2p,\qquad B^-(p)=0,\qquad B^+(q)=1,\qquad B^-(q)=2q-1.
\]
So
\[
W=1+\eta_q-\gamma_p\in\{0,1,2\}.
\]
Therefore \(\P{W=-1}=\P{W=-2}=0\), and the desired inequalities are immediate.
\end{proof}

\begin{lemma}[Multi-pair bias]
\label{lem:convolution}
Let \(X_1,\dots,X_n\) be independent integer-valued random variables such that $\P{X_i=t}\ge r^t\cd \P{X_i=-t}$ for all $i$ and $t\ge 1$. Then their sum \(D:=X_1+\cdots+X_n\) satisfies $\P{D=t}\ge r^t\cd \P{D=-t}$ for all $t\ge 1$.
\end{lemma}

\begin{proof}
Fix \(t\ge1\). For every integer vector \((x_1,\dots,x_n)\) with
\[
x_1+\cdots+x_n=t,
\]
the reflected vector \((-x_1,\dots,-x_n)\) has sum \(-t\), and
\[
\prod_{i=1}^n \P{X_i=x_i}
\ge
\prod_{i=1}^n r^{x_i}\P{X_i=-x_i}
=
r^t\prod_{i=1}^n \P{X_i=-x_i}.
\]
Summing over all vectors with total \(t\) gives
\[
\P{D=t}\ge r^t\cd\P{D=-t}.
\]
\end{proof}

\begin{lemma}[One replacement step]
\label{lem:local-replacement}
Let \(U_1\) and \(U_2\) be independent copies of \(U\), and let \(D\) be an integer-valued random variable such that \(D\), \(U_1\), \(U_2\) and \(W\) are mutually independent. Assume that
\[
\P{D=t}\ge r^t\cd\P{D=-t}
\qquad\text{for every integer }t\ge1.
\]
Then
\[
\E{\sgn(D+W)}\ge \E{\sgn(D+U_1+U_2)}.
\]
\end{lemma}

\begin{proof}
Set
\[
Z:=U_1+U_2,
\qquad
\delta_d:=\E{\sgn(d+W)}-\E{\sgn(d+Z)}.
\]
Then
\[
\E{\sgn(D+W)}-\E{\sgn(D+Z)}
=
\sum_{d\in\mathbb Z}\P{D=d}\cd\delta_d.
\]

Intuitively, because the split ratings $W$ have lower variance than $U_1+U_2$ and satisfy the ratio bound, replacing $U_1+U_2$ by $W$ shifts probability mass from negative to positive outcomes. We consider two cases.

\tbf{Case 1: \(0 \le p < q \le \fr12\) or \(\fr12 \le p < q \le 1\).}
Since $|W|,|Z|$ are bounded by $2$, $d+W$ and $d+Z$ have the same sign as $d$ for \(|d|\ge3\), namely \(\delta_d=0\) for \(|d|\ge3\). Hence, we consider $d \in \{\pm2,\pm1,0\}$. Note that
\[
\delta_{-2}=-a^2,
\qquad
\delta_0=a^2-c^2,
\qquad
\delta_2=c^2,
\]
while
\[
(\delta_{-1},\delta_1)=(-L,L),
\]
where
\[
L=
\begin{cases}
2pq(p+q-pq), & 0<p< q\le \tfrac12,\\[1ex]
2(1-p)(1-q)(1-pq), & \tfrac12\le p< q<1.
\end{cases}
\]
In particular, \(L\ge0\). Therefore
\begin{align*}
\E{\sgn(D+W)}-\E{\sgn(D+Z)}
&=
\P{D=0}(a^2-c^2)
+
L\bigl(\P{D=1}-\P{D=-1}\bigr)\\
&\qquad
+
c^2\P{D=2}-a^2\P{D=-2}.
\end{align*}
By assumption,
\[
\P{D=1}\ge \P{D=-1},
\qquad
\P{D=2}\ge r^2\cd\P{D=-2}.
\]
Since \(a=rc\), we have
\[
c^2\P{D=2}-a^2\P{D=-2}
=
c^2\bigl(\P{D=2}-r^2\P{D=-2}\bigr)\ge0.
\]
Also \(a>c\), so \(a^2-c^2\ge0\). Hence every term is nonnegative, and
\[
\E{\sgn(D+W)}\ge \E{\sgn(D+Z)}.
\]

\tbf{Case 2: \(0<p\le \tfrac12\le q<1\).}
Again \(\delta_d=0\) for \(|d|\ge3\), and a direct calculation gives
\[
\delta_1=2p(1-q)(1-q+pq),
\qquad
\delta_{-1}=-\delta_1,
\qquad
\delta_2=c^2,
\]
\[
\delta_0=(p+q-1)^2+2p(q-p)(1-q)\ge0,
\]
and
\[
r\delta_1+\delta_{-1}=2(q-p)(1-q+pq)\ge0,
\qquad
r^2\delta_2+\delta_{-2}=(1-2p)(2q-1)\ge0.
\]
Therefore
\begin{align*}
\E{\sgn(D+W)}-\E{\sgn(D+Z)}
&=
\P{D=0}\delta_0
+
\P{D=1}\delta_1
+
\P{D=-1}\delta_{-1}\\
&\qquad
+
\P{D=2}\delta_2
+
\P{D=-2}\delta_{-2}.
\end{align*}
Since \(\delta_1,\delta_2\ge0\) and \(D\) satisfies the ratio bound,
\begin{align*}
\P{D=1}\delta_1+\P{D=-1}\delta_{-1}
&\ge
\P{D=-1}\bigl(r\delta_1+\delta_{-1}\bigr)\ge0,\\
\P{D=2}\delta_2+\P{D=-2}\delta_{-2}
&\ge
\P{D=-2}\bigl(r^2\delta_2+\delta_{-2}\bigr)\ge0.
\end{align*}
Together with \(\P{D=0}\delta_0\ge0\), this shows that
\[
\E{\sgn(D+W)}\ge \E{\sgn(D+Z)}.
\]
\end{proof}

Now we use the above three lemmas to prove that $\widetilde K_{2m}(p,q)\ge K_{2m}(p,q)$ for $p< q$, which implies \cref{thm:splitting}.

\begin{proof}[Proof of \cref{thm:splitting}]

For the fixed pair \((\theta_1,\theta_2)\), take independent copies
\(
U_1,\dots,U_{2m}\overset{\mathrm{i.i.d.}}{\sim}U
\)
and
\(
W_1,\dots,W_m\overset{\mathrm{i.i.d.}}{\sim}W.
\)
Then
\[
K_{2m}(p,q)
=
\E{\sgn\!\left(U_1+\cdots+U_{2m}\right)},
\qquad
\widetilde K_{2m}(p,q)
=
\E{\sgn\!\left(W_1+\cdots+W_m\right)}.
\]

For \(j=0,1,\dots,m\), define
\[
S_j:=W_1+\cdots+W_j+U_{2j+1}+\cdots+U_{2m}.
\]
Thus
\[
S_0=U_1+\cdots+U_{2m},
\qquad
S_m=W_1+\cdots+W_m,
\]
and so
\[
K_{2m}(p,q)=\E{\sgn(S_0)},
\qquad
\widetilde K_{2m}(p,q)=\E{\sgn(S_m)}.
\]

Now fix \(j\in\{0,\dots,m-1\}\), and write
\[
D_j:=W_1+\cdots+W_j+U_{2j+3}+\cdots+U_{2m}.
\]
Then
\[
S_j=D_j+U_{2j+1}+U_{2j+2},
\qquad
S_{j+1}=D_j+W_{j+1}.
\]

By \cref{lem:block-ratio}, every summand of \(D_j\) satisfies the ratio
bound
\[
\P{X=t}\ge r^t\cd\P{X=-t}
\qquad (t\ge1).
\]
Hence \cref{lem:convolution} implies that \(D_j\) also satisfies
\[
\P{D_j=t}\ge r^t\cd\P{D_j=-t}
\qquad (t\ge1).
\]
Applying \cref{lem:local-replacement} to \(D_j\), we obtain
\[
\E{\sgn(S_{j+1})}\ge \E{\sgn(S_j)}.
\]
Since this holds for all \(j=0,\dots,m-1\), iterating yields
\[
\widetilde K_{2m}(p,q)
=
\E{\sgn(S_m)}
\ge
\E{\sgn(S_0)}
=
K_{2m}(p,q),
\]
as desired.

\emph{Strictness.} The first replacement step, with $j=0$, already yields a strict gain whenever $0\le p<q\le1$ and $(p,q)\neq(0,1)$. Indeed, in the proof of \cref{lem:local-replacement} every summand is nonnegative, and the summand for $d=0$ equals $\P{D_0=0}\,\delta_0$. Here $D_0=U_3+\cdots+U_{2m}$ satisfies
\[
\P{D_0=0}\ge\P{U=0}^{2m-2} = \big((1-p)(1-q)+pq\big)^{2m-2}>0
\]
unless $(p,q)=(0,1)$, and
\begin{align*}
\delta_0 &= a^2-c^2 = (q-p)\big(q(1-p)+p(1-q)\big) > 0 &&\text{in Case 1},\\
\delta_0 &= (p+q-1)^2+2p(q-p)(1-q) > 0 &&\text{in Case 2},
\end{align*}
again unless $(p,q)=(0,1)$. In Case 2, $\delta_0=0$ forces $p+q=1$ and $p(q-p)(1-q)=0$, hence $p=0$ and $q=1$. Therefore $\widetilde K_{2m}(p,q)>K_{2m}(p,q)$ for all such pairs. On the other hand, $\widetilde K_{2m}(p,q)=K_{2m}(p,q)$ when $p=q$, since both sides vanish by symmetry, and when $(p,q)=(0,1)$, since both sides equal $1$. Integrating against $w(\theta_1,\theta_2)$, the increase in $W_k$ is strict if and only if $w>0$ on a set of positive measure of pairs $\theta_1<\theta_2$ with $\beta(\theta_1)<\beta(\theta_2)$ and $(\beta(\theta_1),\beta(\theta_2))\neq(0,1)$. If $w>0$ almost everywhere, this fails only in two cases. Either $\beta$ is almost everywhere constant, so that every pair has $\beta(\theta_1)=\beta(\theta_2)$, or $\beta$ is almost everywhere $\{0,1\}$-valued, that is, a 0-1 step function. This proves the last assertion of \cref{thm:splitting}.
\end{proof}

\subsection{Proof of \Cref{thm:multiple}} \label{a:multiple}

We give the full proof of \Cref{thm:multiple}. Throughout this section, fix the number of ratings $k$ and a weight function $w$, which by \cref{s:model} is nonnegative and integrable on $\{\theta_1<\theta_2\}$. We write $\sgn(x)\in\{-1,0,1\}$ for the sign function. For a pair of qualities $\theta_1<\theta_2$ with scores $n(\theta_1)$ and $n(\theta_2)$, the ranking accuracy is then $\sgn\bp{n(\theta_2)-n(\theta_1)}$.

\paragraph{Setup.}
Any $k$ scales $\beta_1,\lds,\beta_k$ induce, for each quality $\theta$, a random score
\[
    n(\theta)=\sum_{i=1}^k\Bern{\beta_i(\theta)}\in\{0,1,\dots,k\}.
\]
We denote its distribution by $p(\theta) = (p_0(\theta),\lds,p_k(\theta))\in\Delta(\{0,1,\dots,k\})$. Since the ratings of two different items are independent, the platform objective depends on the scales only through the map $\theta\mapsto p(\theta)$:
\begin{equation} \label{eq:multiple-objective}
    W_k(p) = 2\int_{\theta_1<\theta_2} w(\theta_1,\theta_2)\,\sum_{i,j=0}^{k} \sgn(j-i)\, p_i(\theta_1)\,p_j(\theta_2)\,d\theta_1\,d\theta_2 .
\end{equation}
The scales are nondecreasing, and this has an important consequence for the family $\{p(\theta)\}_{\theta\in[0,1]}$: it is \emph{stochastically nondecreasing} in $\theta$. This means that for all $\theta<\theta'$ and all $j$,
\begin{equation} \label{eq:stoch-monotone}
    F_\theta(j) := \P{n(\theta)\le j} \;\ge\; F_{\theta'}(j) .
\end{equation}
To see this, draw independent uniform random variables $U_1,\lds,U_k$ on $[0,1]$ and set $X_i(\theta) := \one\bb{U_i\le\beta_i(\theta)}$. Then $X_i(\theta)\sim\Bern{\beta_i(\theta)}$, so for every $\theta$ the sum
\[
    \sum_{i=1}^k X_i(\theta)
\]
has the law of $n(\theta)$. Since each $\beta_i$ is nondecreasing, $X_i(\theta)\le X_i(\theta')$ for all $i$ whenever $\theta<\theta'$. Therefore
\[
    \sum_{i=1}^k X_i(\theta)\le\sum_{i=1}^k X_i(\theta') \quad\text{almost surely},
\]
which implies \cref{eq:stoch-monotone}.

The proof has four steps. Step 1 relaxes the problem to arbitrary stochastically nondecreasing families and represents each such family through its quantile functions. Step 2 shows that such a family never does better than one of the \emph{deterministic} and \emph{monotone} assignments that it is a mixture of, so that the problem reduces to partitioning $[0,1]$ into $k+1$ ordered intervals. Step 3 chooses the lengths of these intervals, and Step 4 shows that any such partition is implemented by 0-1 step functions. Steps 3 and 4 are reused in \cref{a:adaptive} for the adaptive setting. The families arising there need not be stochastically monotone, so \cref{lem:relaxation} at the end of this section provides a substitute for Steps 1 and 2 that applies to arbitrary families, at the price of an additional assumption on $w$.

\paragraph{Step 1: relaxation and quantile representation.}
We relax the problem and maximize \cref{eq:multiple-objective} over all measurable maps $\theta\mapsto p(\theta)\in\Delta(\{0,1,\dots,k\})$ that satisfy \cref{eq:stoch-monotone}, not only those induced by $k$ nondecreasing scales. The relaxed optimum upper bounds the objective of every design. Fix such a map $p$ and define its \emph{quantile functions}
\[
    Q_\theta(u) := \min\bc{j\in\{0,1,\dots,k\} : F_\theta(j)\ge u}, \qquad u\in(0,1).
\]
We use three standard facts. First, if $U$ is uniform on $(0,1)$, then $Q_\theta(U)$ has the law $p(\theta)$, because
\[
    \P{Q_\theta(U)\le j} = \P{U\le F_\theta(j)} = F_\theta(j).
\]
Second, $Q_\theta(u)$ is nondecreasing in $u$. Third, and most importantly, \cref{eq:stoch-monotone} is equivalent to
\begin{equation} \label{eq:quantile-monotone}
    Q_\theta(u)\le Q_{\theta'}(u) \qquad\text{for all } u\in(0,1) \text{ and all } \theta<\theta'.
\end{equation}
Indeed, $F_\theta\ge F_{\theta'}$ pointwise means that every level $j$ that is admissible for $\theta'$ in the minimum defining $Q_{\theta'}(u)$ is also admissible for $\theta$. We also note that the map $(\theta,u)\mapsto Q_\theta(u)$ is jointly measurable, because $\{Q_\theta(u)\le j\} = \{F_\theta(j)\ge u\}$ and $\theta\mapsto F_\theta(j)$ is measurable.

Now let $U$ and $U'$ be independent uniform random variables on $(0,1)$. For $\theta_1<\theta_2$, the scores $n(\theta_1)$ and $n(\theta_2)$ are independent with laws $p(\theta_1)$ and $p(\theta_2)$. The pair $\bp{Q_{\theta_1}(U), Q_{\theta_2}(U')}$ has the same joint law, so \cref{eq:multiple-objective} can be written as
\begin{align}
    W_k(p) &= 2\int_{\theta_1<\theta_2} w(\theta_1,\theta_2)\,\E{\sgn\bp{Q_{\theta_2}(U')-Q_{\theta_1}(U)}}\,d\theta_1\,d\theta_2 \nonumber\\
    &= \int_0^1\!\!\int_0^1 \Psi(u,u')\,du\,du', \label{eq:objective-quantile}
\end{align}
where
\[
    \Psi(u,u') := 2\int_{\theta_1<\theta_2} w(\theta_1,\theta_2)\,\sgn\bp{Q_{\theta_2}(u')-Q_{\theta_1}(u)}\,d\theta_1\,d\theta_2 .
\]
The exchange of the expectation and the integral is justified by Fubini's theorem, since the integrand is bounded in absolute value by the integrable function $w$.

\paragraph{Step 2: an optimal family is deterministic and monotone.}
For each $u\in(0,1)$, consider the \emph{deterministic} assignment $n^{(u)}:[0,1]\to\{0,1,\dots,k\}$ defined by $n^{(u)}(\theta) := Q_\theta(u)$. For a deterministic assignment $n$, write its objective as
\[
    V(n) := 2\int_{\theta_1<\theta_2} w(\theta_1,\theta_2)\,\sgn\bp{n(\theta_2)-n(\theta_1)}\,d\theta_1\,d\theta_2 .
\]
Then $V(n) = W_k(\delta_n)$, where $\delta_n$ denotes the family of point masses $\theta\mapsto\delta_{n(\theta)}$, and $\Psi(u,u) = V(n^{(u)})$. By \cref{eq:quantile-monotone}, every $n^{(u)}$ is \emph{nondecreasing} in $\theta$. We claim that for all $u,u'\in(0,1)$,
\begin{equation} \label{eq:symmetrization}
    \Psi(u,u') + \Psi(u',u) \;\le\; V\bp{n^{(u)}} + V\bp{n^{(u')}} .
\end{equation}
Since $w\ge0$, it suffices to prove the corresponding pointwise inequality for the integrands. That is, for every $\theta_1<\theta_2$,
\begin{equation} \label{eq:pointwise}
    \sgn(b'-a) + \sgn(b-a') \;\le\; \sgn(b-a) + \sgn(b'-a'),
\end{equation}
where $a := Q_{\theta_1}(u)$, $b := Q_{\theta_2}(u)$, $a' := Q_{\theta_1}(u')$ and $b' := Q_{\theta_2}(u')$. By symmetry we may assume $u\le u'$. Then $a\le a'$ and $b\le b'$ because quantile functions are nondecreasing in $u$, and $a\le b$ and $a'\le b'$ by \cref{eq:quantile-monotone}. In particular, the right side of \cref{eq:pointwise} is a sum of two terms in $\{0,1\}$. We distinguish three cases according to the second term on the left.
\begin{itemize}[leftmargin=*]
    \item If $b>a'$, then the left side is at most $2$. Since $b>a'\ge a$ and $b'\ge b>a'$, the right side equals $1+1=2$.
    \item If $b=a'$, then the left side is at most $1$. If the right side were $0$, we would have $a=b$ and $a'=b'$, hence $a=b=a'=b'$, and the left side would be $0$ as well. Otherwise the right side is at least $1$.
    \item If $b<a'$, then $\sgn(b-a')=-1$ and the left side is at most $0$, while the right side is at least $0$.
\end{itemize}
This proves \cref{eq:pointwise}, and integrating against $w(\theta_1,\theta_2)\ge0$ proves \cref{eq:symmetrization}. Averaging \cref{eq:symmetrization} over independent uniform $u,u'$ and using \cref{eq:objective-quantile} gives
\begin{align*}
    W_k(p) &= \frac12\int_0^1\!\!\int_0^1 \big(\Psi(u,u')+\Psi(u',u)\big)\,du\,du' \\
    &\le \frac12\int_0^1\!\!\int_0^1 \big(V(n^{(u)})+V(n^{(u')})\big)\,du\,du' \\
    &= \int_0^1 V\bp{n^{(u)}}\,du \\
    &\le \sup_{u\in(0,1)} V\bp{n^{(u)}} .
\end{align*}
Every $n^{(u)}$ is a nondecreasing map from $[0,1]$ to $\{0,1,\dots,k\}$. In other words, it partitions $[0,1]$ into $k+1$ ordered intervals $I_0,I_1,\lds,I_k$, listed from left to right, where $I_j = \{\theta : n^{(u)}(\theta)=j\}$ and some intervals may be empty. We have therefore shown that
\begin{equation} \label{eq:relaxed-optimum}
    \sup_{p \text{ satisfying } \cref{eq:stoch-monotone}} W_k(p) \;=\; \sup_{I_0,\lds,I_k} W_k(I_0,\lds,I_k),
\end{equation}
where the right side ranges over partitions of $[0,1]$ into $k+1$ ordered intervals. The inequality ``$\ge$'' holds because the family of point masses of such a partition satisfies \cref{eq:stoch-monotone}. Two qualities in different intervals are ranked correctly and two qualities in the same interval tie, so the objective of a partition is
\begin{equation} \label{eq:general-w}
    W_k(I_0,\lds,I_k) = \bar w - 2\sum_{j=0}^{k}\int_{(I_j\times I_j)\cap\{\theta_1<\theta_2\}} w(\theta_1,\theta_2)\,d\theta_1\,d\theta_2,
\end{equation}
where
\[
    \bar w := 2\int_{\theta_1<\theta_2} w(\theta_1,\theta_2)\,d\theta_1\,d\theta_2
\]
is the total weight. Thus the optimal partition minimizes the total weight of pairs that fall in the same interval. The right side of \cref{eq:relaxed-optimum} is the maximum of a continuous function of the boundaries $0\le t_1\le\cdots\le t_k\le1$ over a compact set, where continuity follows from the absolute continuity of the integral of $w$. Hence the supremum is attained.

\paragraph{Step 3: the optimal interval lengths.}
By \cref{eq:relaxed-optimum,eq:general-w}, it remains to choose the lengths $x_0,\lds,x_k\ge0$ of the ordered intervals, subject to
\[
    \sum_{j=0}^k x_j=1 .
\]
Only this step depends on the specific weight function. For Kendall's $\tau$ function, that is, $w\equiv1$, the weight of the pairs inside an interval of length $x$ is $x^2/2$ and $\bar w = 1$. Then \cref{eq:general-w} reads
\[
    W_k(I_0,\lds,I_k) = 1-\sum_{j=0}^{k} x_j^2 = 2\sum_{0\le i<j\le k} x_i x_j .
\]
We maximize
\[
    f(x_0,\lds,x_k) := \sum_{0\le i<j\le k}x_ix_j
\]
subject to the constraint above. By Lagrange multipliers, at an interior optimum the partial derivatives
\[
    \pd{f}{x_j} = \sum_{i\neq j} x_i = 1-x_j
\]
are all equal, which forces $x_0=x_1=\cds=x_k=\fr{1}{k+1}$. No boundary point, at which some $x_j=0$, can be optimal. Indeed, moving a length $\varepsilon$ from the longest interval, of length $x_{\max}$, into an empty one changes $\sum_j x_j^2$ by
\[
    (x_{\max}-\varepsilon)^2+\varepsilon^2-x_{\max}^2 = -2\varepsilon(x_{\max}-\varepsilon)<0
\]
for small $\varepsilon>0$, which increases $f$. Equivalently, the Cauchy--Schwarz inequality gives
\[
    \sum_{j=0}^k x_j^2\ge\fr{1}{k+1},
\]
with equality if and only if all $x_j$ are equal. The optimal intervals are therefore $I_j = [\fr{j}{k+1},\fr{j+1}{k+1})$, with boundaries $\tau_i = \fr{i}{k+1}$ for $i\in[k]$, and the optimal value is $W^\te{multiple}_k = 1-\fr{1}{k+1}$.

For a general weight function, the optimal boundaries $0\le\tau_1\le\cdots\le\tau_k\le1$ are the minimizers of the within-interval weight in \cref{eq:general-w}. \Cref{a:adaptive-spacing} works out two examples. When $w>0$ almost everywhere, the optimal intervals are all nonempty, and hence $0<\tau_1<\cdots<\tau_k<1$. Indeed, if some interval were empty, splitting a nonempty interval into two would strictly increase \cref{eq:general-w}, because the pairs straddling the new boundary would then be ranked correctly instead of tying.

\paragraph{Step 4: implementation by 0-1 step functions.}
Finally, we show that every partition of $[0,1]$ into ordered intervals $I_0,\lds,I_k$ is realized by $k$ rating scales, so that the relaxed optimum \cref{eq:relaxed-optimum} is attained by an actual design. Let the boundaries of the partition be $\tau_1\le\cdots\le\tau_k$, so that $I_j = [\tau_j,\tau_{j+1})$ with $\tau_0:=0$ and $\tau_{k+1}:=1$. For $i\in[k]$, let
\[
    \beta^*_i(\theta) := \one\bb{\theta\ge\tau_i}
\]
be the 0-1 step function with threshold $\tau_i$. It is nondecreasing, hence an admissible scale. If $\theta\in I_j$, then $\tau_i\le\theta$ exactly for the indices $i\le j$. Hence exactly the scales $\beta^*_1,\lds,\beta^*_j$ output a positive rating, and the item receives $n(\theta)=j$ positive ratings deterministically. The scales $(\beta^*_i)_{i\in[k]}$ therefore induce the assignment $\theta\mapsto j$ on $I_j$, and their objective equals $W_k(I_0,\lds,I_k)$. Combined with Steps 1--3, this proves \cref{thm:multiple}. For Kendall's $\tau$ function, the thresholds are $\tau_i=\fr{i}{k+1}$. \qed

\subsection{Generalizing \Cref{thm:multiple} to arbitrary score families}

Step 2 in \Cref{a:multiple} relied on the family $\{p(\theta)\}$ being stochastically nondecreasing, which is guaranteed by the monotonicity of the scales. In the adaptive setting of \cref{s:adaptive}, the platform ranks items by an outcome in a finite set $\Omega$ according to an order $\prec$ of its choice. The resulting family of outcome distributions need not be monotone with respect to $\prec$. For instance, the number of positive responses along a path of the tree is not a monotone function of the quality in general. The following lemma shows that for weight functions such as Kendall's $\tau$, the conclusion of Steps 1 and 2 holds for \emph{every} family, monotone or not.

\begin{lemma}[Relaxation to ordered intervals] \label{lem:relaxation}
    Let $S$ be a finite set with a total order $\prec$. For $s,s'\in S$, write $\sgn_\prec(s,s') := 1$ if $s\prec s'$, $\sgn_\prec(s,s') := -1$ if $s'\prec s$, and $\sgn_\prec(s,s') := 0$ if $s=s'$. Let $w$ be a weight function that is nonincreasing in $\theta_1$ and nondecreasing in $\theta_2$ on $\{\theta_1<\theta_2\}$. Then for every measurable map $\theta\mapsto p(\theta)\in\Delta(S)$,
    \[
        2\int_{\theta_1<\theta_2} w(\theta_1,\theta_2)\sum_{s,s'\in S}\sgn_\prec(s,s')\,p_s(\theta_1)\,p_{s'}(\theta_2)\,d\theta_1\,d\theta_2
        \;\le\; \sup_{I_0,\lds,I_{|S|-1}} W(I_0,\lds,I_{|S|-1}),
    \]
    where the supremum is over partitions of $[0,1]$ into $|S|$ ordered intervals and $W(I_0,\lds,I_{|S|-1})$ is given by \cref{eq:general-w} with $k+1$ replaced by $|S|$. In particular, the right side does not depend on the order $\prec$, and an optimal design never benefits from randomizing over orders.
\end{lemma}

\begin{proof}
    Relabel the elements of $S$ as $0\prec1\prec\cdots\prec|S|-1$, so that $\sgn_\prec(s,s')=\sgn(s'-s)$. Let $K:=|S|-1$. The left side is then \cref{eq:multiple-objective} with the score set $\{0,1,\dots,k\}$ replaced by $\{0,1,\dots,K\}$, and we continue to write $W_k(p)$ for it. The proof replaces the quantile coupling of Step 2 by a discretization, a convexity argument and an exchange argument.

    \emph{Reduction to finitely many qualities.} Fix $N\ge1$ and split $[0,1]$ into the cells $C_\alpha := [\tfrac{\alpha-1}{N},\tfrac{\alpha}{N})$ for $\alpha\in[N]$. Let $p^{(N)}$ be the map that is constant on each cell and equal there to the average of $p$ over the cell. By the Lebesgue differentiation theorem, $p^{(N)}(\theta)\to p(\theta)$ for almost every $\theta$. All entries of $p$ and $p^{(N)}$ lie in $[0,1]$ and $w$ is integrable, so dominated convergence applied to \cref{eq:multiple-objective} gives $W_k(p^{(N)})\to W_k(p)$. Hence it suffices to prove the lemma for maps $p$ that are constant on the $N$ cells, for every $N$. Such a map is a point $(p^1,\lds,p^N)$ in the product of $N$ simplices, where $p^\alpha\in\Delta(\{0,1,\dots,K\})$ is the value of $p$ on $C_\alpha$.

    \emph{Multilinearity and reduction to deterministic maps.} For such a map, \cref{eq:multiple-objective} splits into pairs of qualities in different cells and pairs in the same cell:
    \begin{align*}
        W_k(p) &= 2\sum_{\alpha<\beta}\Big(\int_{C_\alpha\times C_\beta} w\Big)\sum_{i,j}\sgn(j-i)\,p^\alpha_i p^\beta_j \\
        &\qquad + 2\sum_{\alpha}\Big(\int_{(C_\alpha\times C_\alpha)\cap\{\theta_1<\theta_2\}} w\Big)\sum_{i,j}\sgn(j-i)\,p^\alpha_i p^\alpha_j .
    \end{align*}
    The second sum vanishes, because by antisymmetry of $\sgn(j-i)$ every distribution $p$ satisfies
    \[
        \sum_{i,j}\sgn(j-i)\,p_i p_j = 0 .
    \]
    Hence $W_k(p)$ is a multilinear function of $(p^1,\lds,p^N)$: fixing all $p^\beta$ with $\beta\neq\alpha$, it is an affine function of $p^\alpha$. An affine function on a simplex attains its maximum at a vertex. Therefore, for $\alpha=1,2,\lds,N$ in turn, we may replace $p^\alpha$ by a point mass $\delta_{n_\alpha}$ without decreasing $W_k$, where
    \[
        n_\alpha\in\argmax_{i}\, W_k(p^1,\lds,p^{\alpha-1},\delta_i,p^{\alpha+1},\lds,p^N).
    \]
    After $N$ replacements, the map is a deterministic sequence $(n_1,\lds,n_N)\in\{0,1,\dots,K\}^N$ whose objective
    \[
        V(n_1,\lds,n_N) = 2\sum_{\alpha<\beta}\Big(\int_{C_\alpha\times C_\beta}w\Big)\,\sgn(n_\beta-n_\alpha)
    \]
    is at least $W_k(p)$. The same convexity argument shows that randomizing the order $\prec$ cannot help. If the platform ranked $s$ above $s'$ with some probability, the objective would be affine in that probability, so one of the two deterministic choices is at least as good.

    \emph{Exchange argument.} Call a pair of cells $\alpha<\beta$ with $n_\alpha>n_\beta$ an \emph{inversion}. Among the finitely many sequences that maximize $V$ over $\{0,1,\dots,K\}^N$, choose one with the fewest inversions. We show that it has none. Suppose for contradiction that $\alpha<\beta$ is an inversion, with $a:=n_\alpha > b:=n_\beta$. Consider the sequence obtained by \emph{swapping} the two values, $n'_\alpha := b$, $n'_\beta := a$, and $n'_\gamma := n_\gamma$ otherwise. We compare the contributions to $V$ before and after the swap, pair by pair. Only pairs involving $\alpha$ or $\beta$ change.
    \begin{itemize}[leftmargin=*]
        \item \emph{The pair $(\alpha,\beta)$.} Its contribution changes from
        \[
            -2\int_{C_\alpha\times C_\beta}w \quad\text{to}\quad +2\int_{C_\alpha\times C_\beta}w,
        \]
        so it weakly increases.
        \item \emph{A cell $\gamma$ with $\alpha<\gamma<\beta$, with value $c:=n_\gamma$.} The pairs $(\alpha,\gamma)$ and $(\gamma,\beta)$ contribute
        \[
            2\int_{C_\alpha\times C_\gamma}w\cdot\sgn(c-a) + 2\int_{C_\gamma\times C_\beta}w\cdot\sgn(b-c)
        \]
        before the swap and
        \[
            2\int_{C_\alpha\times C_\gamma}w\cdot\sgn(c-b) + 2\int_{C_\gamma\times C_\beta}w\cdot\sgn(a-c)
        \]
        after. Since $b<a$, we have $\sgn(c-b)\ge\sgn(c-a)$ and $\sgn(a-c)\ge\sgn(b-c)$, so each term weakly increases. This case uses only $w\ge0$.
        \item \emph{A cell $\gamma$ with $\gamma>\beta$, with value $c$.} The pairs $(\alpha,\gamma)$ and $(\beta,\gamma)$ contribute
        \[
            2\int_{C_\alpha\times C_\gamma}w\cdot\sgn(c-a)+2\int_{C_\beta\times C_\gamma}w\cdot\sgn(c-b)
        \]
        before the swap and
        \[
            2\int_{C_\alpha\times C_\gamma}w\cdot\sgn(c-b)+2\int_{C_\beta\times C_\gamma}w\cdot\sgn(c-a)
        \]
        after, so the change equals
        \[
            2\Big(\int_{C_\alpha\times C_\gamma}w-\int_{C_\beta\times C_\gamma}w\Big)\big(\sgn(c-b)-\sgn(c-a)\big)\;\ge\;0 .
        \]
        Here $\sgn(c-b)\ge\sgn(c-a)$ because $b<a$. The first factor is nonnegative because $w$ is nonincreasing in its first argument. Indeed, for $\theta_1\in C_\alpha$, $\theta_2\in C_\beta$ and $\theta\in C_\gamma$ we have $\theta_1<\theta_2<\theta$, and hence $w(\theta_1,\theta)\ge w(\theta_2,\theta)$. Integrating over $\theta\in C_\gamma$ and then over $\theta_1\in C_\alpha$ and $\theta_2\in C_\beta$, which have the same length, gives
        \[
            \int_{C_\alpha\times C_\gamma}w\ge\int_{C_\beta\times C_\gamma}w .
        \]
        \item \emph{A cell $\gamma$ with $\gamma<\alpha$, with value $c$.} Symmetrically, the change equals
        \[
            2\Big(\int_{C_\gamma\times C_\beta}w-\int_{C_\gamma\times C_\alpha}w\Big)\big(\sgn(a-c)-\sgn(b-c)\big)\;\ge\;0,
        \]
        using $\sgn(a-c)\ge\sgn(b-c)$ and that $w$ is nondecreasing in its second argument.
    \end{itemize}
    Hence the swap does not decrease the objective, so the swapped sequence is also optimal. It has strictly fewer inversions, for three reasons. First, the pair $(\alpha,\beta)$ is no longer inverted. Second, a cell $\gamma$ between $\alpha$ and $\beta$ with $n_\gamma\in[b,a]$ loses every inversion it had with $\alpha$ or $\beta$ and gains none, since after the swap it sees the value $b\le n_\gamma$ on its left and $a\ge n_\gamma$ on its right. Third, a cell between them with $n_\gamma\notin[b,a]$, and every cell outside the range from $\alpha$ to $\beta$, keeps the same number of inversions with $\{\alpha,\beta\}$, because the values $a$ and $b$ merely exchange places on the same side of $\gamma$. This contradicts the choice of a maximizer with the fewest inversions.

    \emph{Conclusion.} We have shown that for every $N$ and every cell-constant $p$ there is a nondecreasing sequence $(n_1,\lds,n_N)$ whose objective is at least $W_k(p)$. Such a sequence is a partition of $[0,1]$ into $K+1$ ordered intervals with boundaries at multiples of $1/N$. Letting $N\to\infty$ proves the inequality of the lemma. The right side is the maximum of \cref{eq:general-w} over partitions, which involves the order only through the number $|S|$ of intervals. This proves the last assertion.
\end{proof}

\begin{remark}[On the monotonicity assumption in \cref{lem:relaxation}] \label{rem:monotone}
    The third and fourth cases of the exchange argument are the only places that use the assumption that $w$ is nonincreasing in $\theta_1$ and nondecreasing in $\theta_2$. Without it, \cref{lem:relaxation} is false. For example, let $|S|=2$, partition $[0,1]$ into four quarters $A<B<C<D$, and let $w=1$ on $A\times B$ and on $C\times D$ and $w=0$ elsewhere. Consider the non-monotone deterministic assignment with $n = 0$ on $A\cup C$ and $n=1$ on $B\cup D$. It ranks every pair in $A\times B$ and in $C\times D$ correctly and achieves $2\cdot\big(\tfrac1{16}+\tfrac1{16}\big) = \tfrac14$. In contrast, a partition into two ordered intervals ranks at most one of the two groups of pairs correctly and achieves at most $\tfrac18$. This does not contradict \cref{thm:multiple}: the assignment above is not stochastically nondecreasing, so by \cref{eq:stoch-monotone} it cannot be induced by nondecreasing scales. Indeed, for $k=1$, the objective of a nondecreasing scale $\beta$ under this $w$ is
    \[
        \fr18\big[(\bar\beta_B-\bar\beta_A)+(\bar\beta_D-\bar\beta_C)\big]\le\fr18,
    \]
    where $\bar\beta_X$ denotes the average of $\beta$ over $X$. The assumption of \cref{lem:relaxation} holds for Kendall's $\tau$ function, for Spearman's $\rho$ function, and more generally whenever $w$ depends only on the gap $\theta_2-\theta_1$ through a nondecreasing function, or only on $\theta_2$ through a nondecreasing function. Whether \cref{thm:adaptive} remains true for weight functions that violate it is an open question.
\end{remark}

\subsection{Proof of \Cref{thm:adaptive}} \label{a:adaptive}

We give the full proof of \Cref{thm:adaptive}. Fix the number of ratings $k$ and a weight function $w$ that is nonincreasing in $\theta_1$ and nondecreasing in $\theta_2$, as in the statement of \cref{thm:adaptive}. Recall from \cref{s:adaptive} that an adaptive design consists of two components: a complete binary tree $T$ of depth $k$ whose internal nodes are labeled by nondecreasing scales, and a total order $\prec$ on the set of outcomes $\Omega=\{0,1\}^k$. Under $T$, an item of quality $\theta$ traverses the tree from the root. At a node labeled by scale $\beta$, it moves to the child labeled $1$ with probability $\beta(\theta)$ and to the child labeled $0$ otherwise, independently of all previous responses. The item's outcome $\omega(\theta)\in\Omega$ is the leaf it reaches, and we denote its distribution by $p(\theta)=(p_\omega(\theta))_{\omega\in\Omega}\in\Delta(\Omega)$. With $\sgn_\prec$ as in \cref{lem:relaxation}, the platform objective is
\begin{equation} \label{eq:adaptive-objective}
    W_k(T,\prec) = 2\int_{\theta_1<\theta_2} w(\theta_1,\theta_2)\sum_{\omega_1,\omega_2\in\Omega}\sgn_\prec(\omega_1,\omega_2)\,p_{\omega_1}(\theta_1)\,p_{\omega_2}(\theta_2)\,d\theta_1\,d\theta_2 .
\end{equation}
This is \cref{eq:multiple-objective} with the score set $\{0,1,\dots,k\}$, ordered by the usual order on integers, replaced by the outcome set $\Omega$ ordered by $\prec$. We follow the structure of the proof of \cref{thm:multiple} in \cref{a:multiple}.

\paragraph{Why the quantile argument does not apply directly.}
The family $\{p(\theta)\}$ induced by a tree of nondecreasing scales is stochastically nondecreasing with respect to the \emph{lexicographic} order on $\Omega$. To see this, couple the responses of two items through common uniform random variables $U_1,\lds,U_k$ as in \cref{a:multiple}. The item of higher quality receives a response at least as large at the root, and if the two responses are equal, both items proceed to the same child. By induction, the sequence of the higher-quality item is lexicographically at least as large. However, the platform may rank outcomes by an arbitrary order $\prec$, with respect to which the family need not be monotone. Step 2 of \cref{a:multiple} is therefore not available. We use \cref{lem:relaxation} instead, which is why \cref{thm:adaptive} is stated under the monotonicity assumption on $w$.

\paragraph{Steps 1 and 2: reduction to ordered intervals.}
Apply \cref{lem:relaxation} with $S=\Omega$ and the order $\prec$. For every tree $T$,
\[
    W_k(T,\prec) \le \sup_{I_0,\lds,I_{2^k-1}} W(I_0,\lds,I_{2^k-1}),
\]
where the supremum is over partitions of $[0,1]$ into $2^k$ ordered intervals and $W(I_0,\lds,I_{2^k-1})$ is given by \cref{eq:general-w} with $2^k$ intervals. By the last assertion of the lemma, the right side is the same for every order $\prec$, and randomizing over orders does not help either. Hence the maximum platform objective over all adaptive designs is at most the best objective of a partition into $2^k$ ordered intervals. In such a partition, all qualities in the $\ell$-th interval are assigned a common outcome $\omega^{(\ell)}$, where $\omega^{(0)}\prec\omega^{(1)}\prec\cdots\prec\omega^{(2^k-1)}$.

\paragraph{Step 3: interval lengths.}
Step 3 of \cref{a:multiple} applies with $2^k$ intervals in place of $k+1$. For Kendall's $\tau$ function, the objective of ordered intervals of lengths $x_0,\lds,x_{2^k-1}$ is
\[
    1-\sum_{\ell=0}^{2^k-1} x_\ell^2,
\]
which is maximized when all $x_\ell = 2^{-k}$, giving the value $1-2^{-k}$. For a general weight function satisfying the assumption of the theorem, the interval boundaries minimize the within-interval weight in \cref{eq:general-w}. The supremum is attained by the same compactness argument as in \cref{a:multiple}.

\paragraph{Step 4: implementation by binary search.}
It remains to show that any partition of $[0,1]$ into $2^k$ ordered intervals $I_0,\dots,I_{2^k-1}$ is realized by a depth-$k$ tree of nondecreasing rating scales together with a total order on $\Omega$. Let the boundaries of the partition be $0 = t_0 \le t_1\le\cdots\le t_{2^k} = 1$, so that $I_\ell = [t_\ell, t_{\ell+1})$. This is the only step that differs substantially from the non-adaptive setting, where a single layer of $k$ step functions suffices.

Identify each outcome $\omega=(r_1,\dots,r_k)\in\{0,1\}^k$ with the integer
\[
    \sum_{j=1}^k r_j 2^{k-j}\in\{0,1,\dots,2^k-1\},
\]
which is its value as a binary number whose most significant digit is $r_1$. Let $\prec$ be the induced order, which is the lexicographic order on response sequences. Consider a node at level $j\in[k]$ of the tree, reached by the responses $(r_1,\dots,r_{j-1})$ of the first $j-1$ users, and let
\[
    m:=\sum_{i=1}^{j-1} r_i 2^{k-i}.
\]
We say that the intervals $I_m, I_{m+1},\dots,I_{m+2^{k-j+1}-1}$ are \emph{consistent} with this node. They are the $2^{k-j+1}$ intervals whose index, written in binary with $k$ digits, begins with $r_1,\dots,r_{j-1}$. We label the node with the 0-1 step function
\[
    \beta_{(r_1,\dots,r_{j-1})}(\theta) := \one\big[\theta \ge t_{m+2^{k-j}}\big],
\]
whose threshold is the boundary between the lower half $I_m,\dots,I_{m+2^{k-j}-1}$ and the upper half $I_{m+2^{k-j}},\dots,I_{m+2^{k-j+1}-1}$ of the consistent intervals. Each such scale is nondecreasing, hence admissible.

We claim that for every $j\in[k+1]$, an item with quality $\theta\in I_\ell$ reaches, after $j-1$ responses, the node whose consistent intervals contain $I_\ell$. This holds for $j=1$, since all intervals are consistent with the root. Suppose that it holds for $j$. At the node reached, the item is asked the question with threshold $t_{m+2^{k-j}}$. Since $I_\ell$ is one of the consistent intervals, $\theta\ge t_{m+2^{k-j}}$ if and only if $I_\ell$ lies in the upper half. Thus the response $r_j$ is $1$ exactly when the $j$-th binary digit of $\ell$ is $1$, and the item proceeds to the child whose consistent intervals are the half containing $I_\ell$. This proves the claim. For $j=k+1$, the only interval consistent with a leaf is the one indexed by its label, so the claim says that the item reaches the leaf $\omega=\ell$. Hence $\omega(\theta)=\ell$ deterministically for all $\theta\in I_\ell$. The tree therefore performs a binary search for the interval containing $\theta$. It realizes the ordered-interval assignment from Step 2 with the outcomes in lexicographic order, so the relaxed optimum is attained by an admissible adaptive design.

For Kendall's $\tau$ function, the intervals have equal length $2^{-k}$ and $t_\ell = \ell\,2^{-k}$. The binary search therefore uses threshold $\fr12$ at the root and thresholds $\fr14$ and $\fr34$ at level $2$. In general, the node reached by responses $(r_1,\dots,r_{j-1})$ uses the threshold
\[
    t_{m+2^{k-j}} = \sum_{i=1}^{j-1} r_i2^{-i}+2^{-j}.
\]
The outcome $\omega(\theta)$ then consists of the first $k$ binary digits of $\theta$, and $W^\te{adaptive}_k = 1-2^{-k}$. \qed

\subsection{Partial adaptivity} \label{a:partial-adaptive}

\begin{proposition}[Batch-adaptive scales] \label{prop:batch}
    Let $w$ be Kendall's $\tau$ function and let $b$ divide $k$. Suppose the $k$ ratings arrive in $k/b$ batches of $b$ ratings each. The $b$ scales in a batch are fixed simultaneously but may depend on the responses in all previous batches. Items are ranked by the sequence $(c_1,\dots,c_{k/b})$ of positive counts in each batch, according to a total order of the platform's choice. Then the maximum platform objective is $1-\fr{1}{(b+1)^{k/b}}$.
\end{proposition}

\begin{proof}
    Each batch count satisfies $c_i\in\{0,1,\dots,b\}$, so the outcome set $S=\{0,1,\dots,b\}^{k/b}$ has $(b+1)^{k/b}$ elements. Kendall's $\tau$ function satisfies the assumption of \cref{lem:relaxation}. Hence, for every batch-adaptive design and every order on $S$, the objective is at most the best objective of a partition of $[0,1]$ into $(b+1)^{k/b}$ ordered intervals. By Step 3 of \cref{a:multiple}, this equals $1-\fr{1}{(b+1)^{k/b}}$ and is attained only by intervals of equal length.

    This value is achieved by a $(b+1)$-ary search, constructed as in Step 4 of the proof of \cref{thm:adaptive} with the binary expansion replaced by the base-$(b+1)$ expansion. Index the equal-length intervals by $\ell\in\{0,1,\dots,(b+1)^{k/b}-1\}$, and identify the outcome $(c_1,\dots,c_{k/b})$ with the integer
    \[
        \sum_{i=1}^{k/b} c_i (b+1)^{k/b-i},
    \]
    ordered as integers. A node at round $i$ reached by the counts $(c_1,\dots,c_{i-1})$ is consistent with the $(b+1)^{k/b-i+1}$ intervals whose index in base $b+1$ begins with these digits. The $b$ scales used in that round are the 0-1 step functions whose thresholds split these consistent intervals into $b+1$ equal parts, exactly as in Step 4 of \cref{a:multiple} applied to the consistent block. An item in the $c$-th part then receives exactly $c$ positive ratings in that round, so the batch count identifies the part containing $\theta$. After $k/b$ rounds, the item reaches the outcome indexed by its interval deterministically. For $b=k$, there is a single round and the construction reduces to \cref{thm:multiple}, giving $W^\te{multiple}_k$. For $b=1$, it is the binary search of \cref{thm:adaptive}, giving $W^\te{adaptive}_k$.
\end{proof}

\subsection{Optimal threshold spacing for other weight functions} \label{a:adaptive-spacing}

By \cref{eq:relaxed-optimum,eq:general-w}, together with \cref{lem:relaxation} in the adaptive setting, the design problem reduces to choosing the boundaries $0=t_0\le t_1\le\cdots\le t_{n}=1$ of $n$ ordered intervals. Here $n=k+1$ in the non-adaptive setting and $n=2^k$ in the adaptive setting. The boundaries should minimize the total weight of the pairs of qualities that fall in the same interval. We give two examples.
\begin{itemize}[leftmargin=*]
    \item \emph{Spearman's $\rho$ function} $w(\theta_1,\theta_2)=\theta_2-\theta_1$. The within-interval weight of an interval of length $x$ is
    \[
        \int_{0<u<v<x}(v-u)\,du\,dv = \frac{x^3}{6},
    \]
    and the total weight is
    \[
        \bar w = 2\int_{\theta_1<\theta_2}(\theta_2-\theta_1)\,d\theta_1\,d\theta_2 = \fr13 .
    \]
    Writing $x_\ell$ for the length of $I_\ell$, \cref{eq:general-w} gives
    \[
        W = \fr13 - \fr13\sum_{\ell=0}^{n-1} x_\ell^3 .
    \]
    By convexity of $x\mapsto x^3$, this is maximized by equal lengths $x_\ell = 1/n$, with value $\fr13\big(1-\fr{1}{n^2}\big)$. Lagrange multipliers give the same conclusion, since they require all $x_\ell^2$ to be equal. Hence under Spearman's $\rho$ function the optimal thresholds are evenly spaced, exactly as under Kendall's $\tau$ function. The two objectives differ only in their value.
    \item \emph{A weight function that only values the top half of the items}, $w(\theta_1,\theta_2) = \one[\theta_2\ge\tfrac12]$, which is constant in $\theta_1$ and nondecreasing in $\theta_2$. An interval contained in $[0,\tfrac12)$ has within-interval weight $0$. An interval $[a,a+x]$ with $a\ge\tfrac12$ has within-interval weight $x^2/2$. An interval $[a,t]$ with $a<\tfrac12<t$ has within-interval weight
    \[
        \int_{1/2}^{t}(\theta_2-a)\,d\theta_2 = \tfrac12\big(t-\tfrac12\big)\big(t+\tfrac12-2a\big) \ge \tfrac12\big(t-\tfrac12\big)^2,
    \]
    which is at least the weight of $[\tfrac12,t]$, because the pairs $(\theta_1,\theta_2)$ in the interval with $\theta_1<\tfrac12\le\theta_2$ tie. Hence it is never worse to place a boundary exactly at $\tfrac12$, and intervals below $\tfrac12$ beyond the first are wasted. The optimal partition therefore uses one interval $[0,\tfrac12)$ and splits $[\tfrac12,1]$ into $n-1$ intervals of equal length $\fr{1}{2(n-1)}$. The thresholds are $\tau_1=\tfrac12$ and $\tau_i = \tfrac12+\fr{i-1}{2(n-1)}$ for $2\le i\le n-1$, evenly spaced within the region that the platform cares about. Since the total weight is
    \[
        \bar w = 2\int_{1/2}^1\theta_2\,d\theta_2 = \fr34,
    \]
    the optimal value is $\fr34-\fr{1}{4(n-1)}$. In particular, the thresholds are not evenly spaced in $[0,1]$.
\end{itemize}

\end{document}